\documentclass[twocolumn]{aastex62}
\pdfoutput=1 
\usepackage{amstext,amsmath}
\usepackage[T1]{fontenc}
\usepackage[figure,figure*]{hypcap}
\usepackage{multirow}
\usepackage{tabularx}
\usepackage{booktabs}

\newcommand{\bB}{{\boldsymbol{B}}}
\newcommand{\bC}{{\boldsymbol{C}}}
\newcommand{\bD}{{\boldsymbol{D}}}
\newcommand{\bT}{{\boldsymbol{T}}}
\newcommand{\bS}{{\boldsymbol{S}}}
\newcommand{\bof}{{\boldsymbol{f}}}
\newcommand{\bF}{{\boldsymbol{F}}}

\newcommand{\bI}{{\boldsymbol{I}}}

\newcommand{\bn}{{\boldsymbol{n}}}
\newcommand{\bj}{{\boldsymbol{j}}}
\newcommand{\bU}{{\boldsymbol{U}}}
\newcommand{\bV}{{\boldsymbol{V}}}
\newcommand{\br}{{\boldsymbol{r}}}
\newcommand{\bx}{{\boldsymbol{x}}}
\newcommand{\bov}{{\boldsymbol{v}}}
\newcommand{\by}{{\boldsymbol{y}}}

\newcommand{\bxi}{{\boldsymbol{\xi}}}
\newcommand{\bdelta}{{\boldsymbol{\delta}}}
\newcommand{\bDelta}{{\boldsymbol{\Delta}}}
\newcommand{\bPhi}{{\boldsymbol{\Phi}}}
\newcommand{\bSigma}{{\boldsymbol{\Sigma}}}
\newcommand{\bLambda}{{\boldsymbol{\Lambda}}}

\newcommand{\btheta}{{\boldsymbol{\theta}}}

\newcommand{\bgamma}{{\boldsymbol{\gamma}}}

\newcommand{\mL}{{\mathcal{L}}}
\newcommand{\mN}{{\mathcal{N}}}
\newcommand{\mP}{{\mathcal{P}}}
\newcommand{\mZ}{{\mathcal{Z}}}
\newcommand{\bmM}{{\boldsymbol{\mathcal{M}}}}
\newcommand{\Cov}{{\text{Cov}}}

\shorttitle{Global 21-cm signal extraction II: Signal constraints}
\shortauthors{Rapetti et al.}

\begin{document}

\title{\MakeUppercase{Global 21-cm Signal Extraction from Foreground and Instrumental Effects II:}

\MakeUppercase{Efficient and Self-Consistent Technique for Constraining Nonlinear Signal Models}}


\author{David Rapetti}
\affiliation{Center for Astrophysics and Space Astronomy, Department of Astrophysical and Planetary Science, University of Colorado, Boulder, CO 80309, USA}
\affiliation{NASA Ames Research Center, Moffett Field, CA 94035, USA}
\affiliation{Universities Space Research Association, Mountain View, CA 94043, USA}

\author{Keith Tauscher}
\affiliation{Center for Astrophysics and Space Astronomy, Department of Astrophysical and Planetary Science, University of Colorado, Boulder, CO 80309, USA}
\affiliation{Department of Physics, University of Colorado, Boulder, CO 80309, USA}

\author{Jordan Mirocha}
\altaffiliation{CITA National Fellow}
\affiliation{McGill University Department of Physics \& McGill Space Institute, 3600 Rue University, Montr\'eal, QC, H3A 2T8}

\author{Jack~O.~Burns}
\affiliation{Center for Astrophysics and Space Astronomy, Department of Astrophysical and Planetary Science, University of Colorado, Boulder, CO 80309, USA}

\email{David.Rapetti@colorado.edu; Keith.Tauscher@colorado.edu}

\begin{abstract}
We present the completion of a data analysis pipeline that self-consistently separates global 21-cm signals from large systematics using a pattern recognition technique. In the first paper of this series, we obtain optimal basis vectors from signal and foreground training sets to linearly fit both components with the minimal number of terms that best extracts the signal given its overlap with the foreground. In this second paper, we utilize the spectral constraints derived in the first paper to calculate the full posterior probability distribution of any signal parameter space of choice.~The spectral fit provides the starting point for a Markov Chain Monte Carlo (MCMC) engine that samples the signal without traversing the foreground parameter space.~At each MCMC step, we marginalize over the weights of all linear foreground modes and suppress those with unimportant variations by applying priors gleaned from the training set.~This method drastically reduces the number of MCMC parameters, augmenting the efficiency of exploration, circumvents the need for selecting a minimal number of foreground modes, and allows the complexity of the foreground model to be greatly increased to simultaneously describe many observed spectra without requiring extra MCMC parameters. Using two nonlinear signal models, one based on EDGES observations and the other on phenomenological frequencies and temperatures of theoretically expected extrema, we demonstrate the success of this methodology by recovering the input parameters from multiple randomly simulated signals at low radio frequencies (10-200~MHz), while rigorously accounting for realistically modeled beam-weighted foregrounds.
\end{abstract}

\section{Introduction}
\label{sec:intro}

Measurements of the sky-averaged (global) 21-cm signal can be utilized to trace the thermal history of the early Universe.~This permits us to investigate both: (i) astrophysical properties of the first populations of stars, galaxies and black holes during Cosmic Dawn---when these first luminous objects formed---and the overall Epoch of Reionization (EoR) driven by energetic photons emitted by those objects, which ultimately ionized the primordial neutral hydrogen (HI) extinguishing its 21-cm spin-flip signal; and (ii) the underlying cosmological model, including potential exotic phenomena---such as dark matter decay, annihilation and interaction with baryons---affecting the cosmic mean temperature during those epochs and particularly the end of the preceding era, the Dark Ages, before astrophysical sources existed.

Recent results from the Experiment to Detect the Global EoR Signature~\citep[EDGES;][]{Bowman:18} show a 78 MHz absorption trough located within the frequency range expected for Cosmic Dawn. However, the amplitude of this trough is about 2-3 times larger than the maximum depth expected from adiabatic cooling due to the cosmic expansion, in the concordance cosmological constant plus cold dark matter model ($\Lambda$CDM). This has generated numerous attempts to explain such an anomaly via excess cooling from non-standard physics, including dark matter particles scattering off baryons~\citep{Barkana:18, Barkana:18b, Fialkov:18, Loeb:18, Berlin:18}. Other possibilities include modifications of the cosmic radio background due to e.g. population III objects or primordial black holes \citep{Feng:18,Ewall-Wice:18,Ewall-Wice:19,Fialkov:19,Mebane:19}.

Unaccounted systematics could alternatively resolve the current discrepancy between observations and theoretical modeling. Using the processed data set released by the EDGES collaboration,\footnote{http://loco.lab.asu.edu/edges/edges-data-release/} \citet{Bradley:2018} showed that the EDGES result could be explained by resonances due to a ground plane artifact, instead of by a signal from the sky. Due to its relevance, this potential systematic is under further investigation by its proposers and the EDGES collaboration. In addition, other concerns about systematics can also be found in the recent literature \citep{Hills:18,Draine:18,Singh:19,Spinelli:19,Sims:19}.

Importantly, other contemporary global 21-cm experiments are working towards verifying these results: Shaped Antenna measurement of the background RAdio Spectrum \citep[SARAS;][]{Patra:2013, Singh:2017}, Sonda Cosmol\'ogica de las Islas para la Detecci\'on de Hidr\'ogeno Neutro \citep[SCI-HI;][]{Voytek:2014}, Zero-spacing Interferometer Measurements of the Background Radio Spectrum \citep[ZEBRA;][]{Mahesh:2014}, Large-aperture Experiment to detect the Dark Ages \citep[LEDA;][]{Bernardi:2015,Bernardi:2016,Price:2018}, Broadband Instrument for Global HydrOgen ReioNisation Signal \citep[BIGHORNS;][]{Sokolowski:2015}, Probing Radio Intensity at high-Z from Marion \citep[PRIzM;][]{Philip:2018}, Radio Experiment for the Analysis of Cosmic Hydrogen~\citep[REACH;][]{Lera:19}, and the Cosmic Twilight Polarimeter \citep[CTP;][]{Nhan:17, Nhan:19}.

Given the impact of the EDGES results and the increasing efforts of the community towards verify them, it is key to also probe the higher redshift, purely cosmological Dark Ages absorption trough of the global 21-cm signal. For this, a space-based mission such as the Dark Ages Polarimeter PathfindER (DAPPER), which is able to collect data at low radio frequencies ($\sim 17-38$ MHz) in the absence of Earth's ionosphere, and to do so with minimal terrestrial radio frequency interference (RFI) in the shadow of the Moon~\citep{Burns:17,Bassett:19}, will be crucial.

Within this context, we are developing a flexible, end-to-end, data analysis pipeline to optimize global 21-cm experiments. In the first paper of this series \citep[][hereafter Paper I]{Tauscher:18}, we analytically calculated constraints on the spectral shapes of simulated 21-cm signals embedded in $ 10^4$-$10^6$ times larger foregrounds by applying a novel technique combining pattern recognition and information criteria (IC).~Furthermore, this work included an innovative experimental design based on rotation-induced foreground polarization that we will continue to employ here in Paper II and discuss further in Paper III (Tauscher et al., in preparation).

In this second paper of the series, we present how we transform the spectral constraints derived in the first step of the pipeline into constraints on nonlinear signal parameters of interest via a Bayesian Markov Chain Monte Carlo (MCMC) analysis. In contrast with previous global 21-cm signal studies, we implement a simultaneous nonlinear fit of signal and foreground by marginalizing over the SVD foreground parameters at each step instead of including them in the parameter space explored by the MCMC. This marginalization, which drastically improves MCMC efficiency, is performed analytically thanks to the foreground model being linear. Including only the signal parameters in the MCMC allows us to combine multiple data spectra, utilize correlations between them and make fits more reliable.

With our pipeline, we show how an MCMC analysis can efficiently find input values of global 21-cm signal parameters in the presence of large beam-weighted foregrounds without a priori knowledge on the region of parameter space in which these values reside. The MCMC is initialized using a mean and covariance derived from a Fisher matrix based procedure that converts the spectral signal constraints of Paper I into estimates of signal parameters for any given model.

In Section~\ref{sec:pipeline}, we sketch the three main components of the full data analysis pipeline:~signal extraction, conversion between parameter spaces, and conditional Bayesian inference. Section~\ref{sec:astro} contains the motivations behind the signal models that we select as examples to test the pipeline under different frequency-dependent forms. Motivated by the EDGES results~\citep{Bowman:18}, we employ two signal models that allow for departures from standard 21-cm shapes.~One is a flattened Gaussian as used by the EDGES team to fit their recent observations (Section~\ref{sec:flattened_gaussian}) and the other a parametric model motivated by theoretical thermal history milestones commonly referred as turning points (Section~\ref{sec:turning-points}).\footnote{We postpone for future work analyses on standard physical models such as those produced by the $\texttt{ares}$ code~\citep[\url{https://bitbucket.org/mirochaj/ares};][]{Mirocha:12, Mirocha:14}, as well as on recently proposed add-on parameterizations to account for excess cooling mechanisms~\citep{Mirocha:19}.} We describe the construction of realistically simulated data from our input signals and foreground modeling in Section~\ref{sec:sims}, and present test cases for both signal models in Section~\ref{sec:results}. We finally discuss further progress to pursue and summarize in Section~\ref{sec:conclusions}.

The Python code, known as $\texttt{pylinex}$, underpinning our analyses in Papers I and II is publicly available.\footnote{\url{https://bitbucket.org/ktausch/pylinex}} This software is particularly useful to fit measurements for which no analytical modeling for the signal and/or the systematics is known and a large covariance between them is present.

\section{SVD/MCMC 21-cm pipeline analysis}
\label{sec:pipeline}

Using simulated data, in Paper I we demonstrated that we can extract a wide variety of 21-cm signals from large foregrounds by fitting Singular Value Decomposition (SVD) eigenmodes derived from two separate training sets, one for the signal and the other for the foreground, and by selecting the number of modes for each set via the Deviance Information Criterion (DIC).\footnote{We selected the DIC over other IC investigated in Paper I, which are also readily available in the \texttt{pylinex} code, because we found it to have the best performance in minimizing signal bias in simulations. However, other IC's could straightforwardly be contributed to that list, and further studies to investigate whether and how, if so, the optimal IC depends on the data analysis and/or experimental setup would certainly be valuable for the community.} We review this procedure in Section~\ref{sec:signal_extraction}.

The next steps of the pipeline are to (i) transform the constraints from the SVD linear fit into a physically-motivated signal parameter space, and (ii) perform a nonlinear MCMC fit of this signal space by conditionalizing over the SVD foreground parameter space. We describe these methods in Sections~\ref{sec:transforming-to-physical-parameters} and~\ref{sec:mcmc_fit}, respectively.

\subsection{First step: signal extraction with \texttt{pylinex}}
\label{sec:signal_extraction}

As a brief summary of Paper I, we remind the reader that we introduced \texttt{pylinex} as a generic scheme to efficiently and rapidly\footnote{The default is a linear analytical calculation, although it can be extended to perform numerical fits.} separate an arbitrary number of distinct sources of information (`data components', in the terminology of Paper I), intrinsically mixed by the experiment (in a fashion described by `expansion matrices' in Paper I), plus random noise, whose properties can be described by either an analytic/numerical framework (such as theoretical modeling of the global signal), simulations (e.g.~radio antenna beam patterns) or lab/sky measurements (e.g.~receiver calibrations/foreground observations).

We tested this separation capability by systematically building realistic data sets combining two of these components---21-cm signals generated with \texttt{ares} and Gaussian beam-weighted foregrounds---with statistical noise whose level is given by the radiometer equation. A natural extension of this initial exercise is to incorporate additional systematics into the foreground training set, such as for instance from a receiver (Paper IV of this series; Tauscher et al., in preparation).

Using \texttt{pylinex}, after defining SVD models for the different data components and combining them into a single linear model, we solve for the coefficients in that model $\bxi$ and their covariance matrix $\bS$ using
\begin{equation}
  \bS = (\bF^T\bC^{-1}\bF)^{-1} \ \ \text{ and } \ \ \bxi = \bS\bF^T\bC^{-1}\by,
\end{equation}
where $\by$ is the data vector, $\bC$ is the covariance matrix of the noise distribution, and $\bF$ is a matrix with the SVD basis vectors as columns. From $\bxi$ and $\bS$, we analytically calculate the maximum likelihood estimate $\gamma_k$ of each data component $\by_k$, its channel covariance, $\bDelta_k$, and its averaged $1\sigma$ root-mean-square error, $\text{RMS}_k$, through
\begin{subequations} 
\begin{align}
  \bgamma_k &= \bF_k \bxi_k~, \label{eq:channel-mean} \\
  \bDelta_k &= \bF_k \bS_{kk} {\bF_k}^T, \label{eq:channel-covariance} \\
  \text{RMS}_k &= \sqrt{\frac{\text{Tr}(\bDelta_k)}{n_k}}\;, \label{eq:channel-rms}
\end{align} 
\end{subequations}

\noindent where $\bxi_k$ is the portion of the parameter mean $\bxi$ containing parameters modeling $\by_k$, $\bS_{kk}$ is the diagonal block of the parameter covariance matrix $\bS$ corresponding to those parameters, and $n_k$ is the number of data channels in the $\bF_k$ basis.

\subsection{Second step: Transforming to physical parameters} \label{sec:transforming-to-physical-parameters}

The first step after obtaining SVD signal parameter distributions from \texttt{pylinex} is to approximately transform them into the chosen space of physically-motivated signal parameters by searching for the best fit in the target parameter space. Na\"ively, one might attempt to do this by fitting the \texttt{pylinex}-outputted signal band in frequency space and minimizing a likelihood such as
  \begin{equation}
    \mL_{\text{LSF-na\"ive}}(\btheta_{21}) \propto \exp{\left\{ -\frac{1}{2} \bdelta_\nu^T \bDelta_{21}^{-1} \bdelta_\nu \right\}}, \label{eq:lsf-naive}
  \end{equation}
  where $\bdelta_\nu\equiv\by_{21}-\bmM_{21}(\btheta_{21})$, $\by_{21}$ is the signal associated with the mean of the SVD coefficient distribution, $\bmM_{21}(\btheta_{21})$ is the physical signal model evaluated at the signal parameter vector $\btheta_{21}$, and $\bDelta_{21}$ is the diagonal matrix whose elements are the variances of the frequency channels under the \texttt{pylinex} fit, as defined in Equation~\ref{eq:channel-covariance}. Minimizing this likelihood corresponds to directly fitting the bands shown in Figure 7 of Paper I. However, for our purpose, this fit is insufficient to start our MCMC sampler for two reasons:
  
  \begin{enumerate}
  \item It does not use all information from \texttt{pylinex} due to the fact that the covariance matrix $\bDelta_{21}$ only accounts for channel variances.~Attempting to account for channel covariances makes $\bDelta_{21}$ singular since there are more frequency channels than there are SVD signal modes.
  \item A single parameter vector, such as the one found by the least square fit, cannot be used to initialize multiple MCMC chains.
  \end{enumerate}

\subsubsection{Transforming signal into SVD parameters}

The first issue above can be solved by performing the least square fit in SVD coefficient space instead of frequency space. This is performed by redefining the likelihood function being minimized as
  \begin{equation}
    \mL_{\text{LSF}}(\btheta_{21}) \propto \exp{\left\{ -\frac{1}{2} \bdelta^T\bS_{21}^{-1}\bdelta \right\}}, \label{eq:lsf-full}
  \end{equation}
  where $\bdelta\equiv\bxi_{21}-\bPhi_{21}\bmM_{21}(\btheta_{21})$ now represents the displacement of the physical signal associated with $\btheta_{21}$ transformed into SVD coefficient space from the mean $\bxi_{21}$ of the SVD coefficient distribution with respect to its covariance $\bS_{21}$. The matrix $\bPhi_{21}$, which transforms a signal in frequency space $\bT_{21}$ to the SVD coefficient vector $\bx_{21}$ that minimizes the weighted least squares residual $(\bF_{21}\bx_{21}-\bT_{21})^T\bC^{-1}(\bF_{21}\bx_{21}-\bT_{21})$, where $\bF_{21}$ is the matrix with signal basis vectors as its columns and $\bC$ is the full data noise covariance matrix, is given by
  \begin{equation}
    \bPhi_{21} = (\bF_{21}^T\bC^{-1}\bF_{21})^{-1}\bF_{21}^T\bC^{-1}. \label{eq:signal-projection-matrix}
  \end{equation}
  Note that $\bdelta=\bPhi_{21}\bdelta_\nu$.~Unlike minimizing the likelihood defined in Equation~\ref{eq:lsf-naive}, minimizing the likelihood in Equation~\ref{eq:lsf-full} builds in all SVD covariances consistently and concisely.~We denote the parameter vector that minimizes the likelihood of Equation~\ref{eq:lsf-full} as $\overline{\btheta_{21}}$.

\subsubsection{Fisher matrix formalism}
\label{sec:fisher_matrix}

To solve the second issue in Section~\ref{sec:transforming-to-physical-parameters}, we estimate the signal parameter covariances by inverting the Fisher information matrix of $\mL_{\text{LSF}}$ (Equation~\ref{eq:lsf-full}), i.e.
  \begin{subequations} \begin{align}
    \Lambda^{(0)}_{ij} &\equiv \Cov[(\btheta_{21})_i,(\btheta_{21})_j] \\ &\approx \left[\left(\bD^T\bS_{21}^{-1}\bD\right)^{-1}\right]_{ij},
  \end{align} \end{subequations}
  where $\bLambda^{(0)}$ is the initial covariance matrix  estimate and
  \begin{equation}
  \bD\equiv\left.\bPhi_{21}\frac{\partial\bmM_{21}}{\partial\btheta_{21}}\right|_{\btheta_{21}=\overline{\btheta_{21}}}.   
  \end{equation}

\noindent In cases where the gradient $\frac{\partial\bmM_{21}}{\partial\btheta_{21}}$ is not implemented or is difficult to compute analytically, it is estimated numerically from appropriately chosen finite steps.
  
Under the Fisher approximation, our initial estimate of the signal distribution in physical parameter space is a multivariate Gaussian given by
\begin{equation}
    \btheta_{21}\sim\mN\left\{\overline{\btheta_{21}},\bLambda^{(0)}\right\}. \label{eq:signal-guess-distribution}
\end{equation}

\subsection{Third step: MCMC fit}
\label{sec:mcmc_fit}

The fit performed through the methods of Section~\ref{sec:transforming-to-physical-parameters} does not include the SVD foreground parameters, whose distribution must be considered when exploring the distribution of signal parameters.~As long as the signal model is nonlinear, this must be calculated with numerical sampling.~For this purpose, we have implemented a custom MCMC sampler in the \texttt{pylinex} code, based upon a Metropolis-Hastings (MH) algorithm \citep{Gelman:14}.~While MH samplers are simple to implement, they have a disadvantage in that a significant amount of information must be supplied up front.~Specifically, one must generate not only a probability density function to sample, but also a proposal distribution, which determines the probability density of a chain moving from a starting to an ending point, and a distribution from which to draw initial points for individual, independent MCMC chains.~We solve these drawbacks as follows in Sections~\ref{sec:mcmc-likelihood},~\ref{sec:guess-distribution},~\ref{sec:proposal-distribution},~and~\ref{sec:updating}.\footnote{A different path for proceeding without a proposal distribution is to use nested sampling, for instance through the widespread \texttt{MultiNest} \citep{Feroz:09} and \texttt{PolyChord} \citep{Handley:15} codes. In this case, one specifies a prior volume which is then whittled down into the final distribution, instead of the distribution being built up via MCMC sampling.~This alternative sampling method could also be readily incorporated into our engine.}

\subsubsection{Probability density function}
\label{sec:mcmc-likelihood}

  A natural choice of the probability density function (PDF) to explore is the likelihood function multiplied by priors,
  \begin{equation}
    p(\btheta_{\text{fg}},\btheta_{21})= \frac{\pi_{\text{fg}}(\btheta_{\text{fg}})\ \pi_{21}(\btheta_{21})\ \mL(\btheta_{\text{fg}}, \btheta_{21})}{\mZ}. \label{eq:full-probability-density}
  \end{equation}
  In this equation, $\pi_X(\btheta_X)$ are the priors on the $X$ parameters, $\mZ$ is a $\btheta$-independent constant, and the likelihood function is
  \begin{multline}
    \mL(\btheta_{\text{fg}}, \btheta_{21}) = |2\pi\bC|^{-1/2} \\ \times \exp{\left\{-\frac{1}{2}[\br(\btheta_{\text{fg}},\btheta_{21})]^T\bC^{-1}[\br(\btheta_{\text{fg}},\btheta_{21})]\right\}},
  \end{multline}
  where $\br(\btheta_{\text{fg}},\btheta_{21})=\by-\bmM_{\text{fg}}(\btheta_{\text{fg}})-\bmM_{21}(\btheta_{21})$ is the residual of the model of the data vector $\by$, written with the foreground and 21-cm components separated.
  
  The PDF given by Equation~\ref{eq:full-probability-density} is the joint density of all parameters, but we aim at only exploring numerically the signal parameters, $\btheta_{21}$.~To do so, we modify the density to be the marginal signal parameter distribution by integrating over $\btheta_{\text{fg}}$, yielding
  \begin{equation}
    p(\btheta_{21}) = \frac{\pi_{21}(\btheta_{21})}{\mZ}\ \int \mL(\btheta_{\text{fg}},\btheta_{21})\ \pi_{\text{fg}}(\btheta_{\text{fg}})\ d\theta_{\text{fg}}. \label{eq:signal-probability-density}
  \end{equation}
  We define $\mP(\btheta_{\text{fg}},\btheta_{21})\equiv\mL(\btheta_{\text{fg}},\btheta_{21})\ \pi_{\text{fg}}(\btheta_{\text{fg}})$ for convenience, which is, up to a multiplicative constant, equal to the conditional posterior PDF of the foreground parameters when the signal parameters are $\btheta_{21}$. Therefore, up to such a multiplicative constant, the integral is equal to the conditional Bayesian evidence of the foreground model when the signal parameters are fixed to $\btheta_{21}$. When the foreground priors are Gaussian, as we take them to be, $\mP$ is Gaussian in $\btheta_{\text{fg}}$ and the integral is equal to $|2\pi\bSigma_{\mP}(\btheta_{21})|^{1/2}\ \mP_{\text{max}}(\btheta_{21})$ where $\mP_{\text{max}}(\btheta_{21})$ is the maximum value of $\mP(\btheta_{\text{fg}}, \btheta_{21})$ with $\btheta_{21}$ fixed and $\bSigma_{\mP}(\btheta_{21})$ is the covariance of the Gaussian form of $\mP(\btheta_{\text{fg}},\btheta_{21})$ with $\btheta_{21}$ fixed.\footnote{When the foreground model is linear and its priors are Gaussian or nonexistent, $\bSigma_{\mP}(\btheta_{21})$ is independent of $\btheta_{21}$.} Plugging in this evaluation of the integral, the marginal signal PDF from Equation~\ref{eq:signal-probability-density} satisfies
  \begin{equation}
    p(\btheta_{21}) \propto \pi_{21}(\btheta_{21})\ \mP_{\text{max}}(\btheta_{21})\ |\bSigma_{\mP}(\btheta_{21})|^{1/2}. \label{eq:final-signal-probability-density}
  \end{equation}
  Sampling this distribution instead of the PDF in Equation~\ref{eq:full-probability-density} is much faster because the dimension of the explored space is greatly reduced, and it uses the knowledge that the conditional distribution of the foreground is Gaussian by analytically marginalizing over foreground parameters at each MCMC step instead of numerically exploring them.

\subsubsection{Initial distribution of MCMC iterates} \label{sec:guess-distribution}

The initial guess distribution for the signal parameters is given by the output of step 2 of the pipeline (Equation~\ref{eq:signal-guess-distribution}), which is normal with mean $\overline{\btheta_{21}}$ and covariance $\bLambda^{(0)}$.

\subsubsection{Proposal distribution}
\label{sec:proposal-distribution}

One of the most critical inputs to an MH MCMC sampler is its proposal distribution, the distribution from which it draws new points at which to evaluate the probability density in Equation~\ref{eq:final-signal-probability-density}. If the variances (diagonal components of the proposal distribution covariance) are too narrow, nearly all steps will be accepted but the sampler will not move efficiently through the parameter space. If they are too broad, nearly all steps will be rejected because the sampler will attempt to move too far in parameter space.

The off-diagonal components of the covariance are also important.~For constant variances, excluding the off-diagonal covariances leads to a $1\sigma$ interval whose hypervolume is $\left|\frac{\text{det}(\bC_{\text{diag}})}{\text{det}(\bC_{\text{full}})}\right|$ times larger than the same interval when the full covariance is used, which, in most cases, leads to a similar situation as mentioned above in the case where the variances are too broad.

\begin{figure*}[tb]
  \centering
  \includegraphics[width=0.49\textwidth]{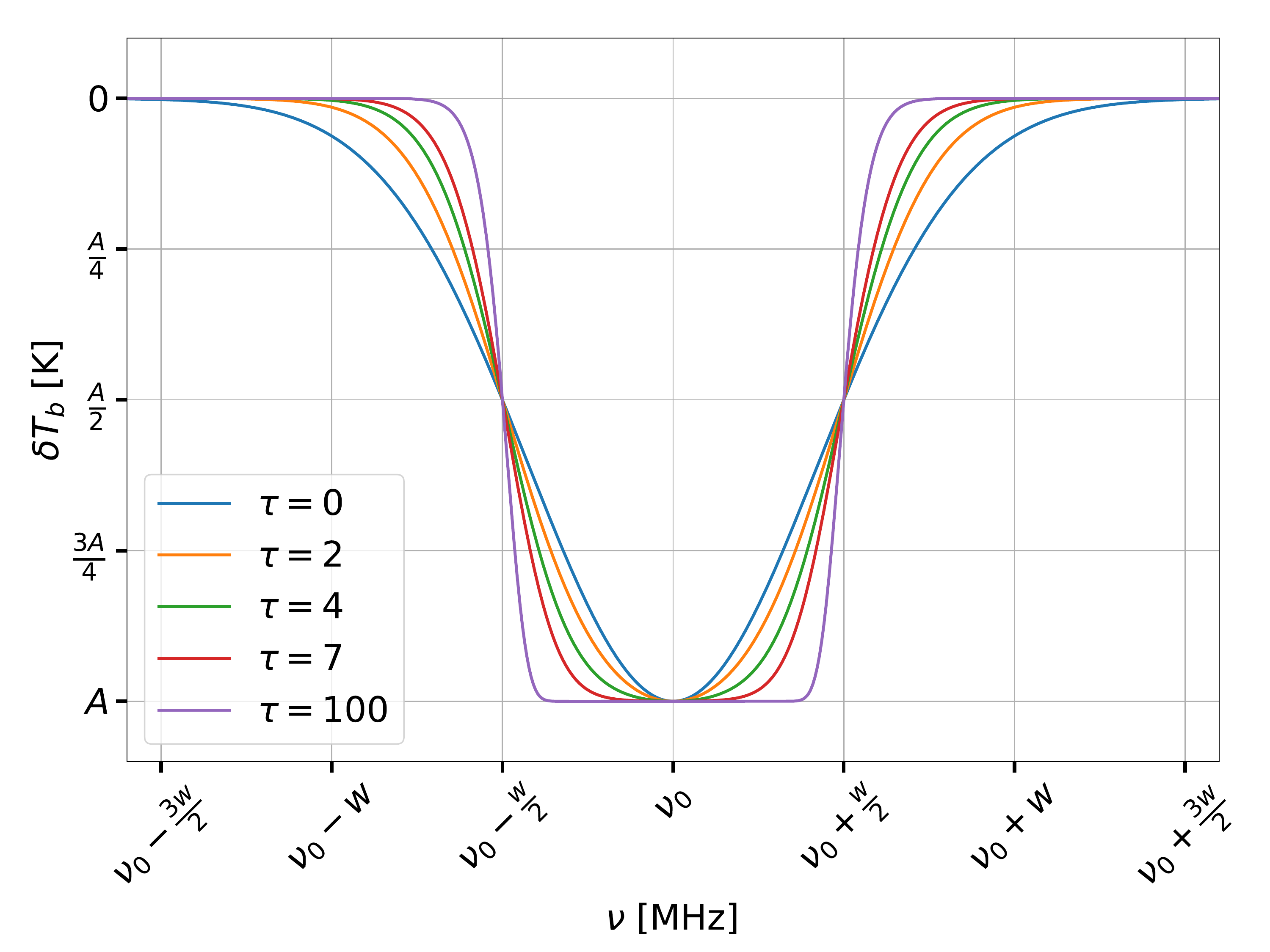}
  \includegraphics[width=0.49\textwidth]{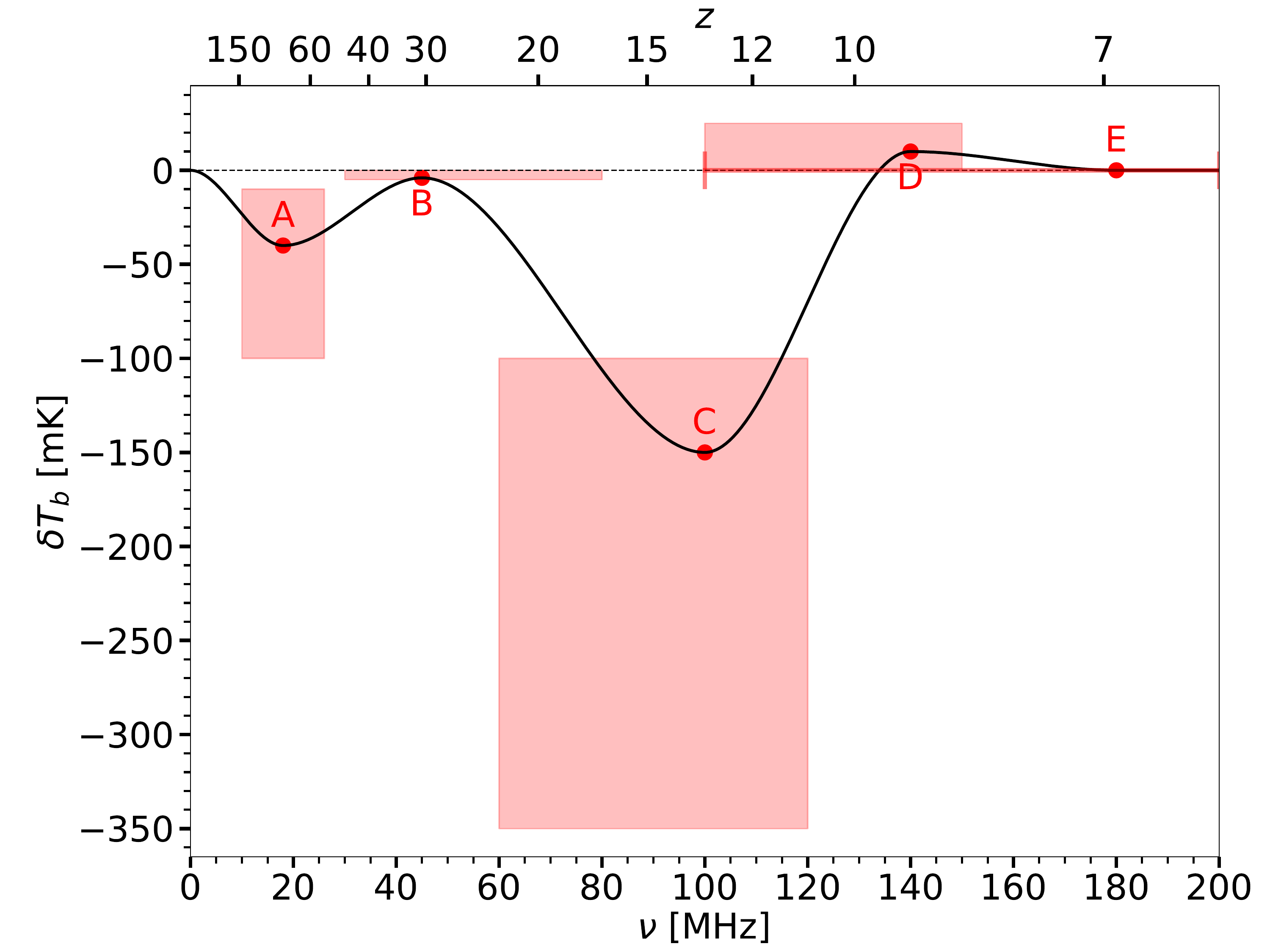}
  \caption{\textit{Left}: Flattened Gaussian model dependence on parameters $A$, $\nu_0$, $w$, and $\tau$. The first three parameters shift and scale the signal while $\tau$ (at constant $w$) determines how long around $\nu_0$ the signal stays near its maximum depth. \textit{Right}: A typical turning point model with the movable points defining the spline interpolation marked by red dots. The spline is also constrained so that $\delta T_b$ and its derivative are 0 at $\nu=0$. Broadly speaking, $A$ represents the Dark Ages, $B$ represents Cosmic Dawn, $C$ through $D$ represents the epoch of heating, and $D$ through $E$ represents the epoch of reionization. The filled regions around turning points $A$-$D$ show the allowed positions of the points. The horizontal line with vertical bars on its ends marks the allowed frequencies of turning point E. In addition to the constraints shown, in any given realization, the frequencies of adjacent turning points are forced to be at least 10 MHz apart.} \label{fig:model-explanations}
\end{figure*}

The covariance matrix of the initial Gaussian proposal distribution for the parameters is equal to $\bLambda^{(0)}/c(\alpha)$ where $\bLambda^{(0)}$ is the covariance matrix of the initial distribution of MCMC iterates and $c(\alpha)$, as defined in Appendix~\ref{app:updating}, is a proportionality constant meant to achieve an acceptance fraction $\alpha$.~Therefore, the probability density of proposing a jump from $\btheta_{21}^{(i)}$ to $\btheta_{21}^{(f)}=\btheta_{21}^{(i)}+\bxi$ is
  \begin{multline}
    p \left(\btheta_{21}^{(i)} \rightarrow \btheta_{21}^{(i)}+\bxi\right) = \sqrt{c(\alpha)}\ \lVert2\pi\bLambda^{(0)}\rVert^{-1/2} \\ \times\ \exp{\left\{ -\frac{c(\alpha)}{2}\bxi^T(\bLambda^{(0)})^{-1}\bxi \right\}}. \label{eq:initial-proposal-distribution}
  \end{multline}

\subsubsection{Updating and acceptance rate}
\label{sec:updating}

In order to increase the efficiency of the MCMC search when using the basic MH algorithm, we schedule updates of the proposal based on the given distributions of all the MCMC chains up that time. At the $k^{\text{th}}$ update, the covariance of recently visited points is computed and denoted $\bLambda^{(k)}$. Then, the proposal matrix is updated to $\bLambda^{(k)}/c(\alpha)$ where, once again, $\alpha$ is the desired acceptance fraction, leading the jumping probability to be equal to that shown in Equation~\ref{eq:initial-proposal-distribution} with $\bLambda^{(0)}$ replaced by $\bLambda^{(k)}$.

\subsection{Foreground priors}
\label{sec:foregroundterms}

In our MCMC fit, we use a very large number of foreground terms. This is sensible because we use the foreground training set to seed prior information.~We fit each curve in the training set with the linear model and a Gaussian approximation of the resulting eigenmode coefficients is computed.\footnote{This can be achieved without any extra computation when the eigenmodes of the curves themselves form the linear model. See Appendix~\ref{app:priors-from-svd}.}~Then, we use Gaussian distributions with the means and variances of the mode weights obtained from these fits as priors. While using the foreground modes themselves relies on the training set variations being similar in form to the data, using these priors amounts to the assumption that the magnitude of the data variations is similar to the magnitudes found in the training set.

\section{Nonlinear 21-cm signal models}
\label{sec:astro}

For testing purposes, we will examine two physically-motivated models. One based on EDGES observations (Section~\ref{sec:flattened_gaussian}) and the other on key physical processes theoretically predicted to govern the time evolution of the global 21-cm signal (Section~\ref{sec:turning-points}).

\subsection{Flattened Gaussian model}
\label{sec:flattened_gaussian}

First we will demonstrate our pipeline using an analytical model that was recently fitted to EDGES data by \cite{Bowman:18}. This is a flattened Gaussian model with four parameters: the amplitude $A$, center frequency $\nu_0$, full width at half maximum $w$, and flattening $\tau$. In terms of these parameters, the 21-cm signal is modeled as
  \begin{multline}
    T_{21}(\nu) = A\ \frac{1-e^{-\tau e^B}}{1-e^{-\tau}} \ \ \\ \text{ where } \ \ B = \left[\frac{\nu-\nu_0}{(w/2)}\right]^2\ \ln{\left[-\frac{1}{\tau}\ln{\left(\frac{1+e^{-\tau}}{2}\right)}\right]}.
  \end{multline}

\noindent This is a phenomenological model with no physical motivation beyond representing an absorption trough.~It was adopted by the EDGES collaboration for its ability to significantly reduce the RMS of the residuals when fitting their data~\citep{Bowman:18}.~For these fits, they used foreground models based on polynomial expansions around the dominant power law behaviour. Two of them, however, were loosely inspired by ionospheric effects, but the parameters obtained where clearly unphysical as pointed out by \cite{Hills:18} \cite[see also the EDGES reply in][]{Bowman:18b}.

Despite the shortcomings of the flattened Gaussian model, its simplicity makes it a useful initial example to exercise our pipeline. The left panel of Figure~\ref{fig:model-explanations} shows how $\nu_0$, $w$, and $A$ shift and scale the model, as well as the effect of the flattening parameter $\tau$, which continuously modulates the shape of the signal between a Gaussian ($\tau\rightarrow 0$) and a square pulse ($\tau\rightarrow\infty$).

\begin{figure*}[tb]
\begin{center}
\includegraphics[width=0.49\textwidth]{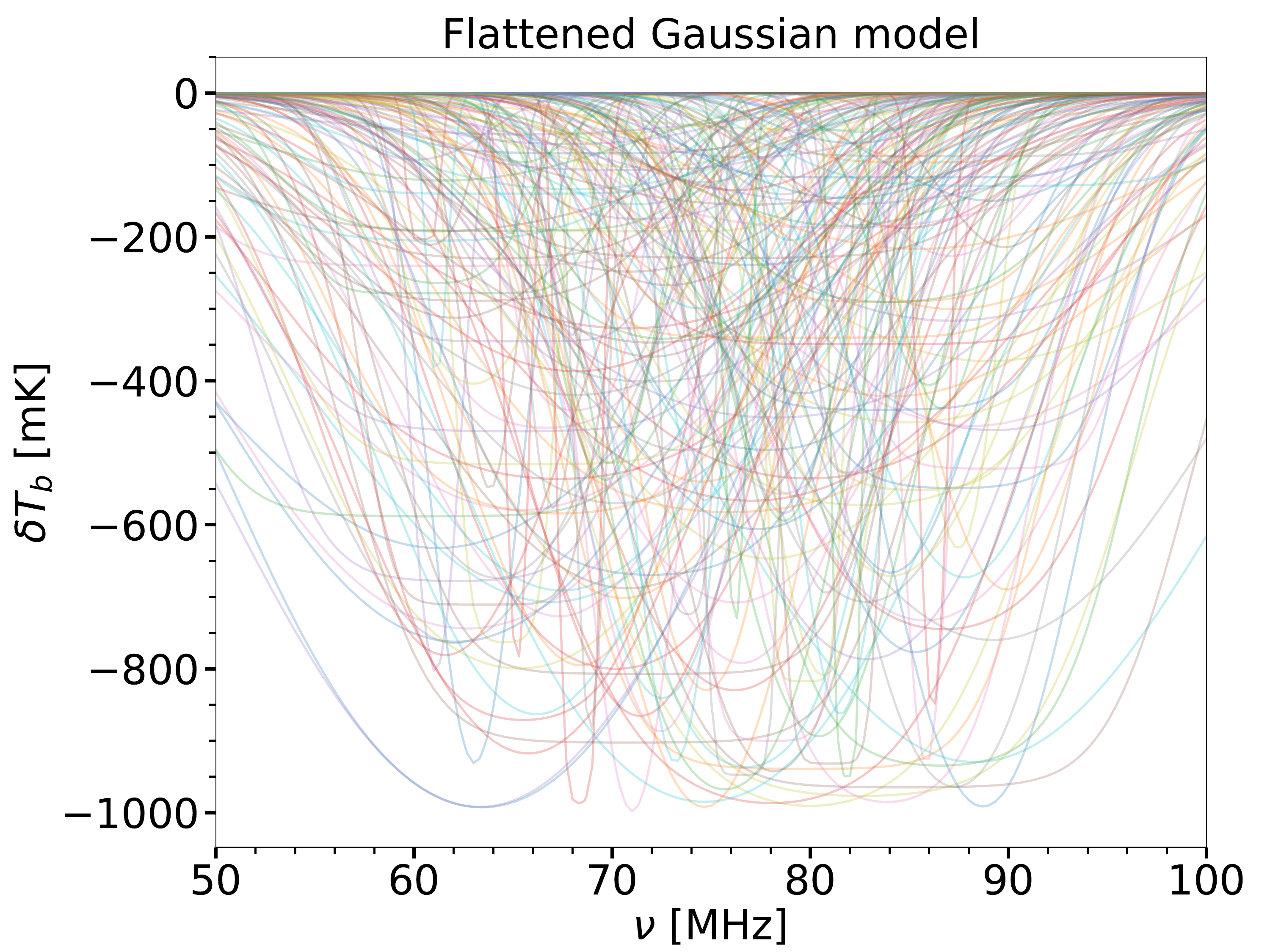}
\includegraphics[width=0.49\textwidth]{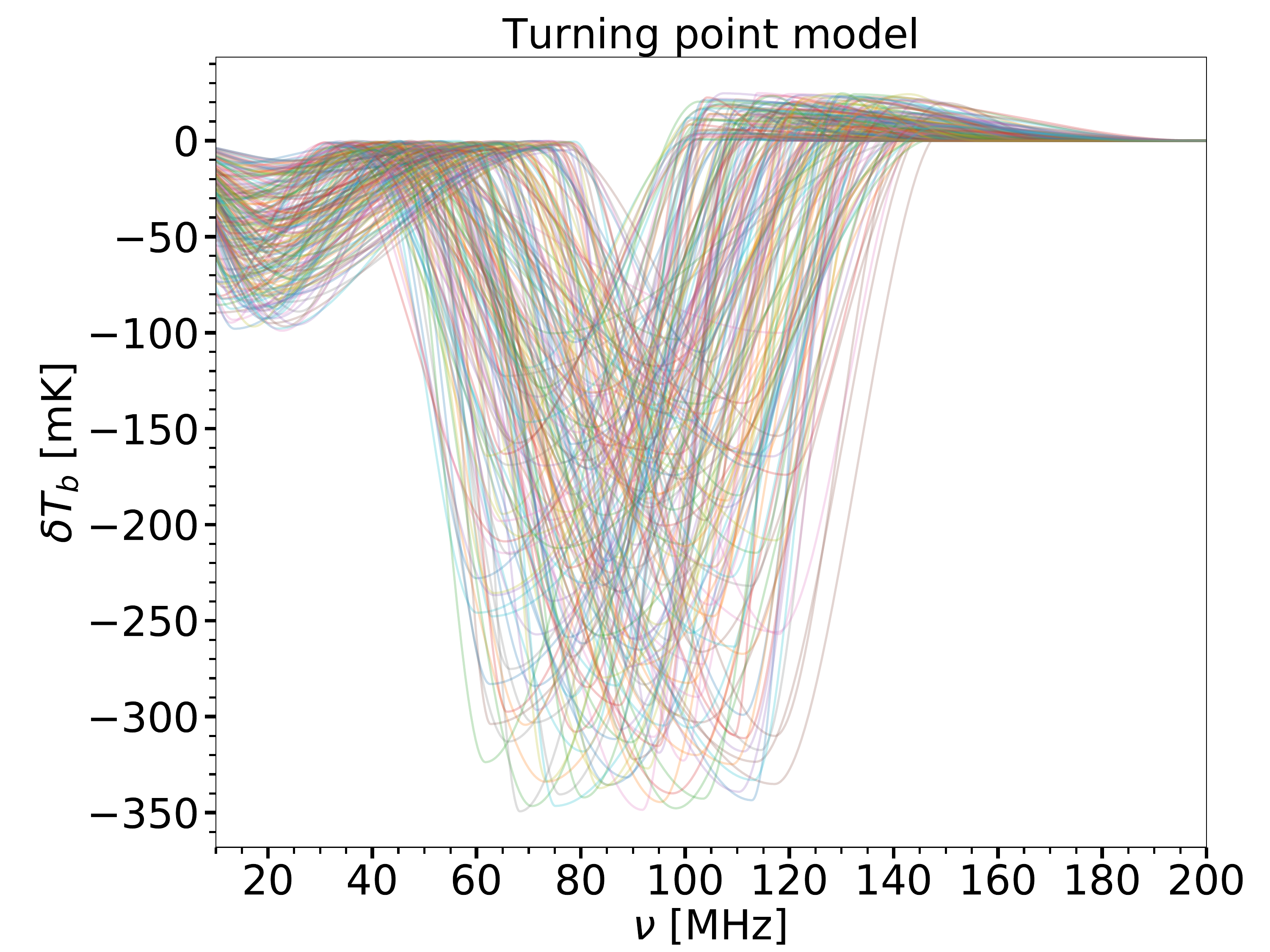}
\caption{\textit{Left}: Sample of 200 curves from the training set for the flattened Gaussian model. Table~\ref{tab:flattened-gaussian-training-set-distribution} describes the distribution of flattened Gaussian parameters in the training set. \textit{Right}: Sample of 200 curves from the training set for the turning point model. Table~\ref{tab:turning-point-training-set-distribution} describes the distribution of turning point frequencies and temperatures in the training set.}
\label{fig:training-sets}
\end{center}
\end{figure*}

\subsection{Turning point model}
\label{sec:turning-points}

Second, we parametrize the global 21-cm signal based on physically motivated extrema in its spectral shape, known as turning points~\citep{Pritchard:10, Harker:16}. These are milestones in the cosmic history of the hydrogen gas.

Briefly, after recombination decoupled the gas from photon temperature, the 21-cm spin temperature coupled to that of the gas.~Since the gas cooled faster than the cosmic microwave background (CMB), the signal, which is the 21-cm brightness temperature relative to the CMB, goes into absorption.~When the coupling of the 21-cm brightness to the gas temperature became ineffective compared to the coupling to the CMB because of the low gas density, the 21-cm temperature recoupled to that of the CMB, causing the signal to turn around and creating an absorption trough with a minimum typically labelled turning point A. At turning point B, the first stars turn on. Via the Wouthuysen-Field effect \citep{Wouthuysen:52,Field:58}, the Lyman-$\alpha$ radiation from the first stars recoupled the 21-cm transition to the temperature of the gas, which had continued cooling with respect to the CMB, triggering another absorption trough.~From its minimum, turning point C, the signal rises back due to the first stars and black holes significantly heating the gas. At turning point D, the reionization of the gas begins extinguishing the signal down to its disappearance at turning point E.

\begin{table}[tb]
    \centering
    \caption{Probability densities of parameters in the flattened Gaussian signal training set} \label{tab:flattened-gaussian-training-set-distribution}
    \begin{tabularx}{\columnwidth}{@{\extracolsep{\stretch{1}}}*{4}{c}@{}}
        \hline
        \hline
        Symbol & Parameter & Units & Distribution \\
        \hline
        $A$ & Amplitude & K & Unif(-1, -0.1) \\
        $\nu_0$ & Center & MHz & Unif(60, 90) \\
        $w$ & FWHM & MHz & Unif(1, 30) \\
        $\tau$ & Flattening & N/A & Exp(1) \\
        \hline
    \end{tabularx}
\end{table}

The free parameters of the model are the frequencies and brightness temperatures of the turning points, except for the temperature of E which is fixed to zero. The model is a cubic spline between the turning points. In order to force these points to be extrema (i.e. have derivative zero), each turning point uses 2 spline knots placed at the same temperature and 20 kHz apart symmetrically around the turning point frequency given by the parameters. In addition to the turning points A to E, there are two knots placed at 0 K and $10\pm10$ kHz. The model always evaluates to 0 K at frequencies above that of turning point E. The right panel of Figure~\ref{fig:model-explanations} shows schematically the relative locations and allowed ranges (red rectangles) for the turning point modeling that we use here.

\begin{table}[tb]
    \centering
    \caption{Probability densities of parameters in the turning point model signal training set}
    \label{tab:turning-point-training-set-distribution}
    \begin{tabularx}{\columnwidth}{@{\extracolsep{\stretch{1}}}*{4}{c}@{}}
        \hline
        \hline
        Symbol & Parameter & Units & Distribution \\
        \hline
        $\nu_A$ & A frequency & MHz & Unif(10, 26) \\
        $T_A$ & A temperature & mK & Unif(-100, -10) \\
        $\nu_B$ & B frequency & MHz & Unif(30, 80) \\
        $T_B$ & B temperature & mK & Unif(-5, 0) \\
        $\nu_C$ & C frequency & MHz & Unif(60, 120) \\
        $T_C$ & C temperature & mK & Unif(-350, -100) \\
        $\nu_D$ & D frequency & MHz & Unif(100, 150) \\
        $T_D$ & D temperature & mK & Unif(0, 25) \\
        $\nu_E$ & E frequency & MHz & Unif(100, 200) \\
        \hline
    \end{tabularx}
    \vspace{0.8ex}
    
    {\raggedright \textbf{Notes.} The frequencies of adjacent turning points are also constrained to differ by at least 10 MHz. \par}
\end{table}

\section{Simulated data}
\label{sec:sims}

\subsection{Signal training sets}
\label{sec:signals}

Signal training sets are formed with the models described above and are seeded with distributions of the underlying parameters. The distributions of the flattened Gaussian parameters are shown in Table~\ref{tab:flattened-gaussian-training-set-distribution}. The training set consists of $10^6$ curves made from parameters drawn from these distributions. The distributions of the turning point parameters are shown in Table~\ref{tab:turning-point-training-set-distribution}. The turning point training set consists of $10^5$ curves drawn from these distributions.~Figure~\ref{fig:training-sets} shows samples of the training sets for both models in frequency space.

\subsection{Foreground modeling}
\label{sec:foreground}

The simulated foregrounds are constructed in the same manner as in Paper I, as briefly described here. For simplicity, we include only beam-weighted foreground emission, ignoring other systematics such as human activity generated RFI, refraction, absorption and emission due to Earth's ionosphere, and receiver gain and noise temperature variations (the latter will be discussed in Paper IV). The experiment simulated here is most analogous to a pair of antennas orbiting the Moon and taking data above the farside, where the ionospheric effects and RFI need not be addressed \citep{Burns:17,Burns:19}.

The simulated data products, $\by$, of all fits in this paper are concatenations of 96 brightness temperature spectra, which include 4 Stokes parameters, $I$, $Q$, $U$, and $V$, at $N_\chi=6$ different rotation angles, $\chi$, about the antenna boresight for $N_{\bn}=4$ different antenna pointing directions, $\bn$. The antenna simulated is a dual-dipole system modulated by angular Gaussian profiles of varying full widths at half maximum.~The sky brightness temperature in the simulations is given by the observed 408 MHz Haslam map \citep{Haslam:82} with each pixel scaled by the power law $[\nu/(408\text{ MHz})]^{-2.5}$. Paper III of this series will explore how measuring Stokes parameters at multiple rotation angles and antenna pointings  makes fits more rigorous and decreases errors. In addition to beam variations, upcoming work will include multiple sky models derived from observations, improving the accuracy of the uncertainties towards realistic analyses.

The variance of the noise added to the data in each fit, $\sigma^2$, is constant across the different Stokes parameters and is related to the total power (Stokes $I$) brightness temperature, $T_b$, through the radiometer equation,
\begin{equation}
    \sigma^2(\nu,\chi,\bn) = \frac{N_{\chi}N_{\bn}}{\Delta \nu\ \Delta t}\ [T_b(\nu,\chi,\bn)]^2,
\end{equation}
with a frequency channel width $\Delta\nu$ of 1 MHz and a total integration time $\Delta t$ of 800 hours.~The data are split into $5$ different components---one for the 21-cm signal and one for the beam-weighted foregrounds (which are correlated across boresight angles and frequency) of each pointing, $\boldsymbol{n}$. The signal is the same across all $N_{\boldsymbol{n}}=4$ pointings while the foregrounds for each pointing only affect the data from that pointing.

\section{Results}
\label{sec:results}

In Section~\ref{sec:extraction_mcmc}, we examine the ability of the pipeline to utilize the SVD spectral constraints on the signal, retrieved without a priori knowledge, to inform our MCMC on the starting location and covariance proposal.~Given a large parameter space such as ours, this initial information is critical for an efficient MCMC search.~We then test two key elements of this analysis in Section~\ref{sec:varying-num-terms}. These are the number of linear SVD foreground terms that will be marginalized over the signal parameter space explored by the MCMC algorithm, and the robustness of this selection given our use of foreground priors (see Section~\ref{sec:foregroundterms} and Appendix~\ref{app:priors-from-svd}). We end the section with an exploration of the constraining power on the Cosmic Dawn and Dark Ages troughs, and how these constraints are affected by statistical noise and systematics confusion (Section~\ref{sec:integration-time}).

\begin{figure}[tb]
  \centering
  \includegraphics[width=0.46\textwidth]{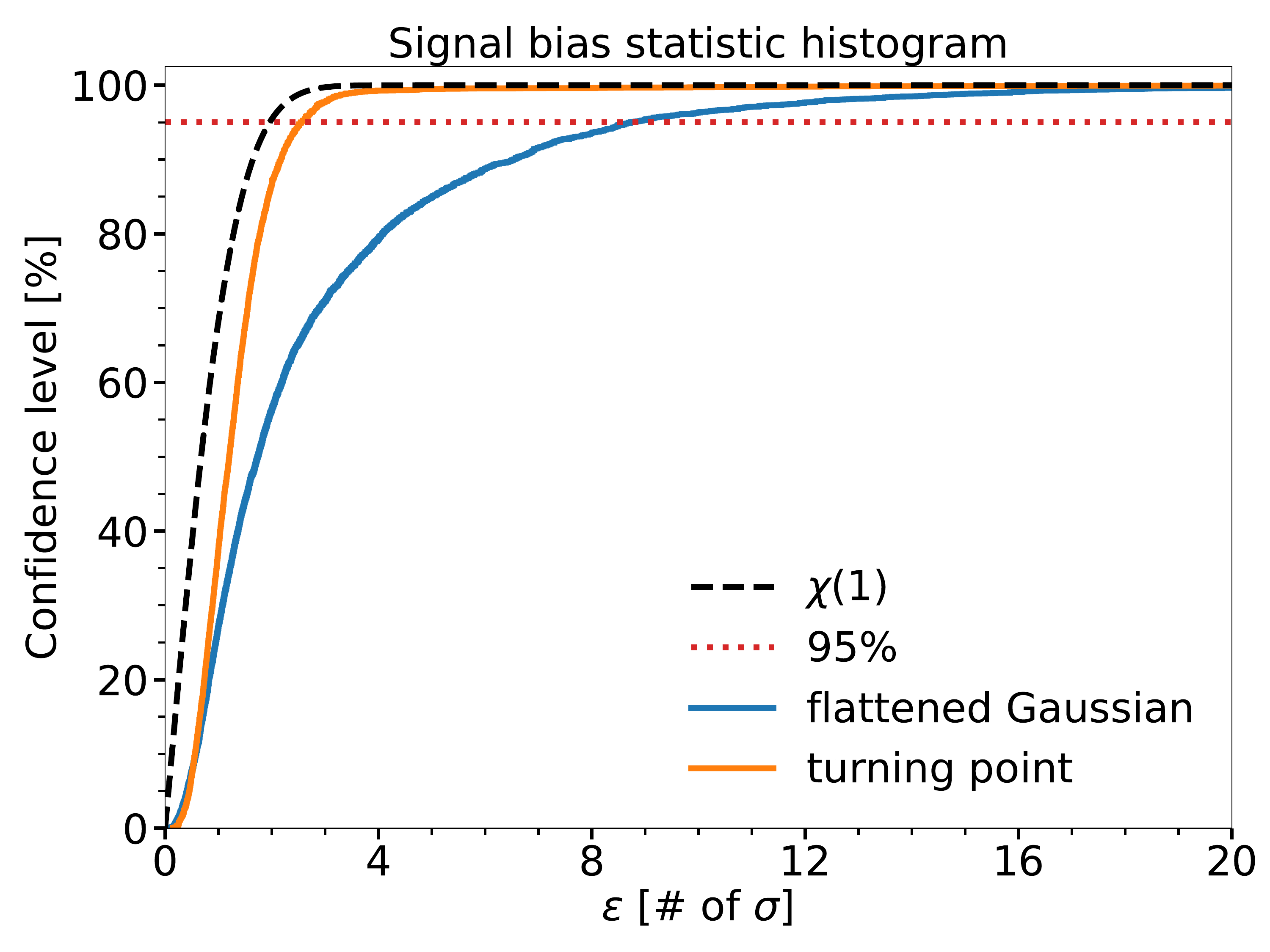}
  \caption{Histogram showing cumulative distributions of the signal bias statistic $\varepsilon$ defined in Paper I (and here recalled in Equation~\ref{eq:signal-bias-statistic}) for the flattened Gaussian (blue curve) and turning point (orange) models. The $\varepsilon$ values where the histograms cross 95\% indicate the $\sigma$ level of the 95\% confidence intervals. The crossing of this value for the two models is very different: for the turning point model the sigma level ($\sim 2.5\sigma$) is relatively close to that of the reference $\chi(1)$ distribution (2$\sigma$), whereas the flattened Gaussian crosses at $\sim 8.75\sigma$ level.}
  \label{fig:linear-calibration}
\end{figure}

\subsection{Signal extraction and MCMC fits}
\label{sec:extraction_mcmc}

\begin{figure}[tb]
\begin{center}
\includegraphics[width=0.47\textwidth]{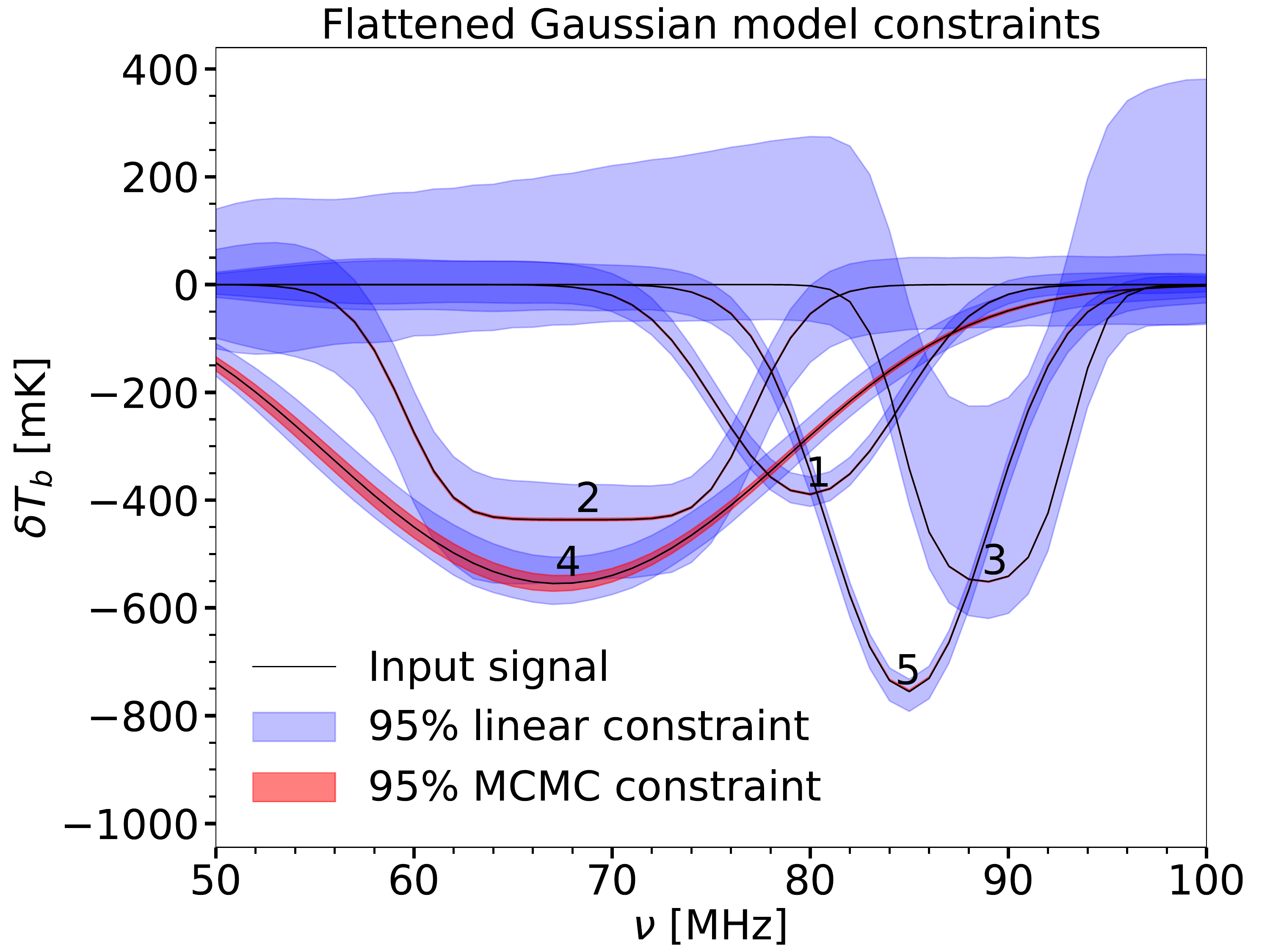}
\caption{
Pipeline constraints in frequency space for the flattened Gaussian model. The blue intervals correspond to the $95\%$ intervals from the linear fit, which uses SVD modes to represent the signal in addition to the foreground, while the red intervals correspond to the $95\%$ confidence intervals from the MCMC fit, which uses the full nonlinear signal model and SVD foreground modes. For the linear fit, the $95\%$ confidence intervals correspond to $8.75\sigma$ (see Figure~\ref{fig:linear-calibration}). The models are denoted as FG1, FG2, FG3, FG4, and FG5 in the text and in Tables~\ref{tab:terms-and-rms}~and~\ref{tab:flattened-gaussian-recovered-parameters}.}
\label{fig:flattened-gaussian-frequency-space}
\end{center}
\end{figure}

We present results for both models of Section~\ref{sec:astro} using various random cases for each. In Section~\ref{sec:refmodel}, we discuss the signal reconstructions of these two sets of cases, achieved first by `linear signal extraction' (initial step of the pipeline as presented in Paper I and summarized in Section~\ref{sec:signal_extraction}) and second by `non-linear conditional Bayesian inference' (third step as presented in Section~\ref{sec:mcmc_fit}). As shown in Figures~\ref{fig:flattened-gaussian-frequency-space}~and~\ref{fig:turning-point-frequency-space}, the excellent matches between the results obtained by each of the two methods serve as proof-of-concepts for these two novel techniques:~(i) pattern recognition based on SVD+IC for the signal extraction and (ii) fitting that marginalizes over linear foreground parameters during an MCMC exploration of a nonlinear signal space. The latter ensures a simultaneous and self-consistent fit of the systematics parameters (in these examples, foreground).

In Section~\ref{sec:mcmcconstraints}, for each model used we present the constraints inferred on their signal parameters, and how we  recover their input values in a statistically consistent manner for all nine random cases fitted, validating the robustness of the pipeline.

\subsubsection{Linear systematic errors}
\label{sec:errors}

As described in Paper I, for the linear signal fits it is critical to calibrate the confidence levels of individual cases, as presented below in Section~\ref{sec:refmodel}, using simulations throughout the corresponding training set.~Figure~\ref{fig:linear-calibration} presents this calibration for both signal models, flattened Gaussian (blue curve) and turning point (orange curve). In addition to the number of parameters selected by the DIC (see Section~\ref{sec:numbers_modes}), the overlap between the SVD signal and foreground modes, which are independently obtained from each training set, is also key in determining the size of the errors. Figure~\ref{fig:linear-calibration} shows that the foreground model has a notably larger overlap with the flattened Gaussian model than with the turning point model. This accentuates the difference between the linear and MCMC fits for each model, as seen when comparing Figures~\ref{fig:flattened-gaussian-frequency-space} and~\ref{fig:turning-point-frequency-space}.

The ideal scenario is having training sets with SVD bases that are orthogonal.~Keeping this in mind when designing an experiment and forming training sets is important. Our use of induced polarization pursues this goal by adding data components to the foreground training set that minimize the overlap with the signal.

\subsubsection{Signal reconstructions}
\label{sec:refmodel}

\begin{figure*}[tb]
\begin{center}
\includegraphics[width=0.48\textwidth]{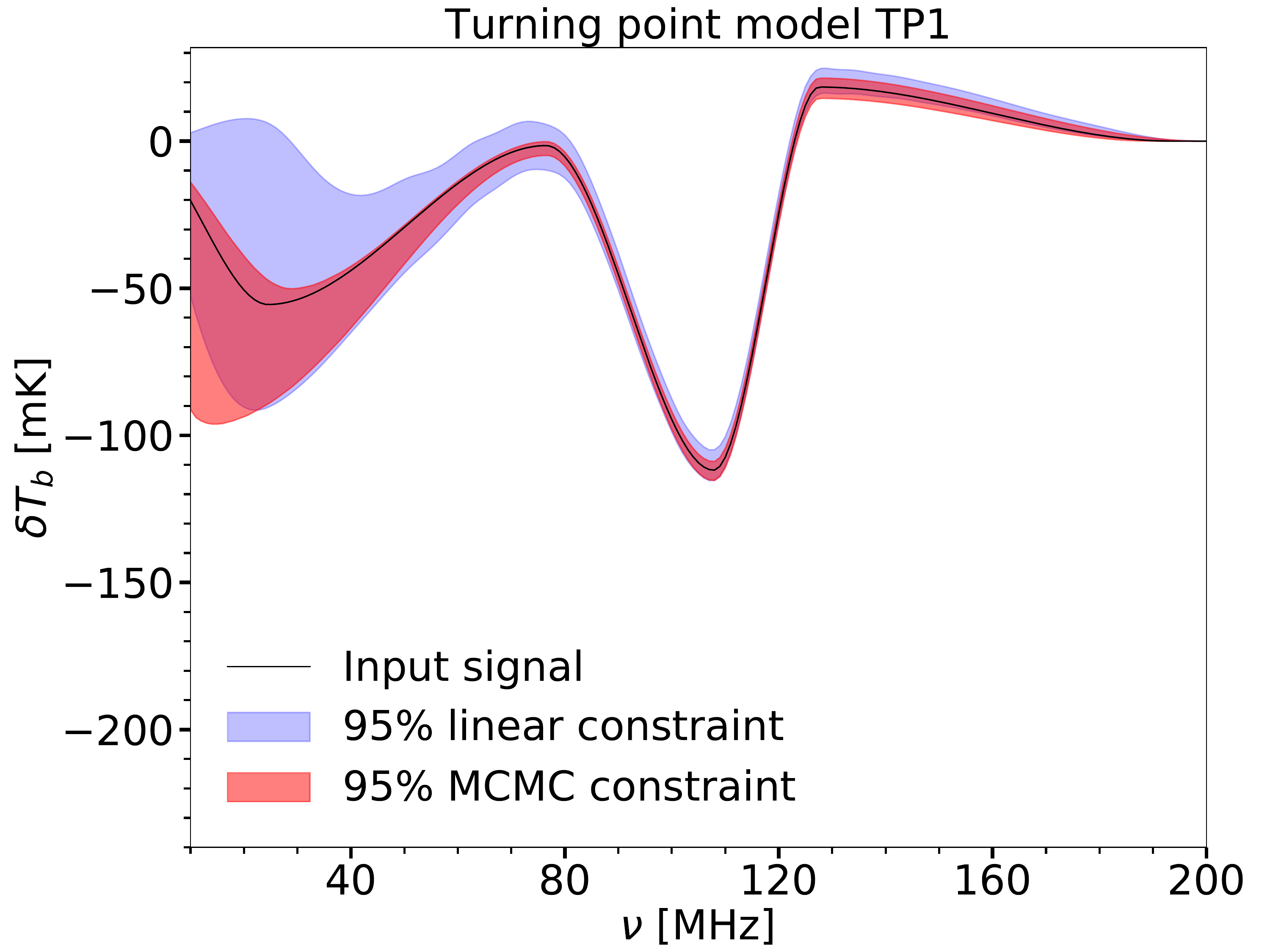}
\includegraphics[width=0.48\textwidth]{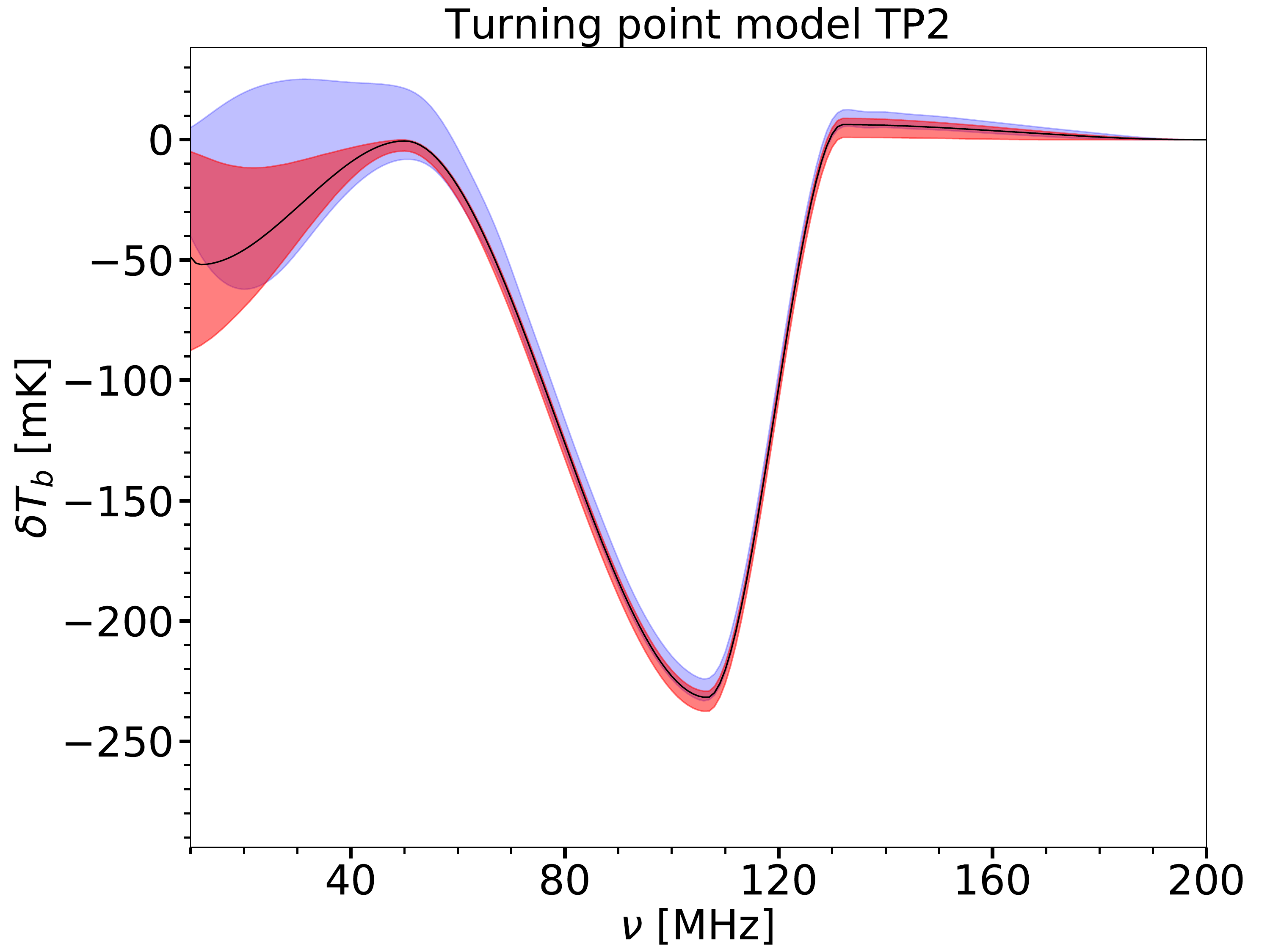}
\includegraphics[width=0.48\textwidth]{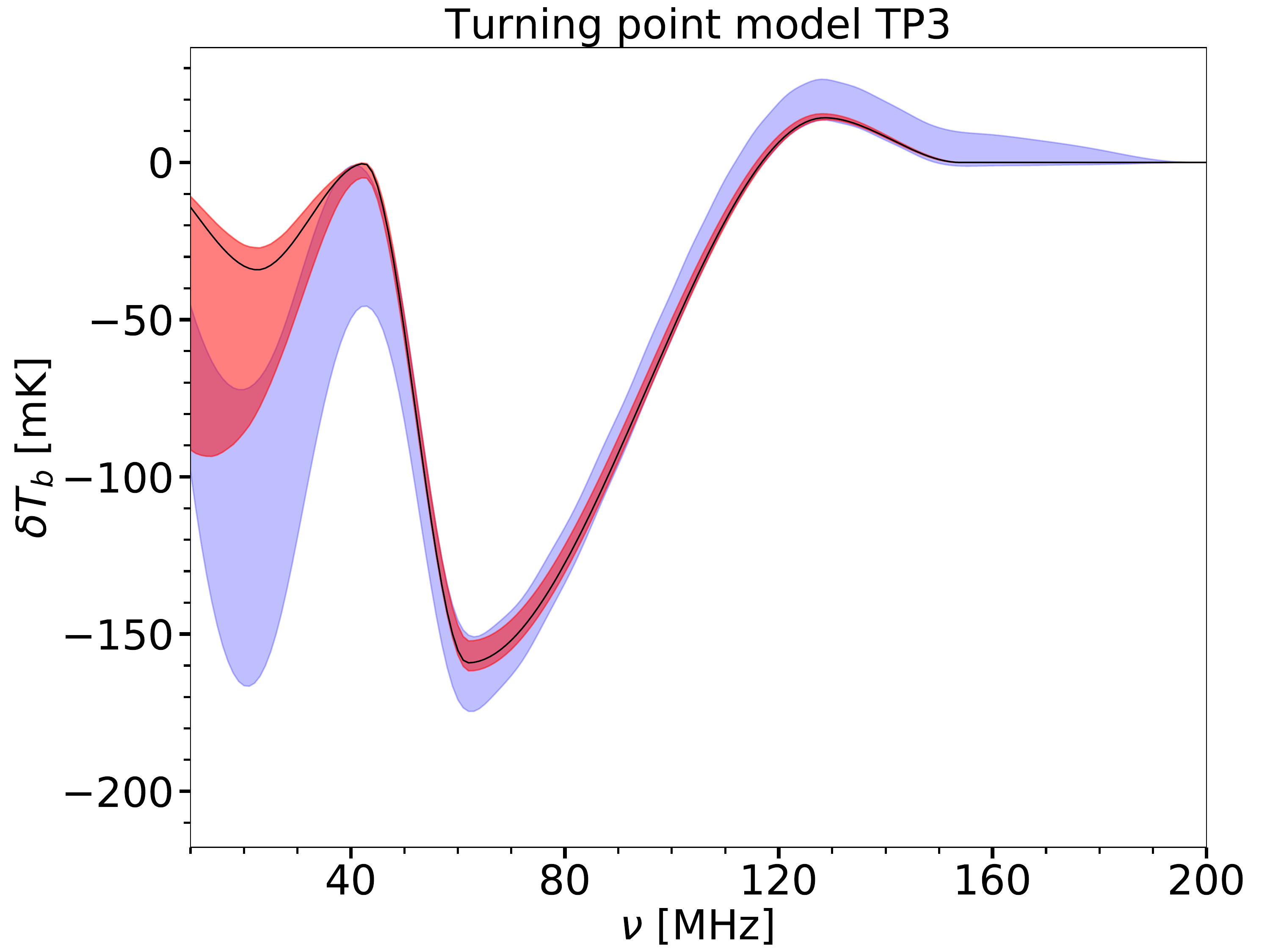}
\includegraphics[width=0.48\textwidth]{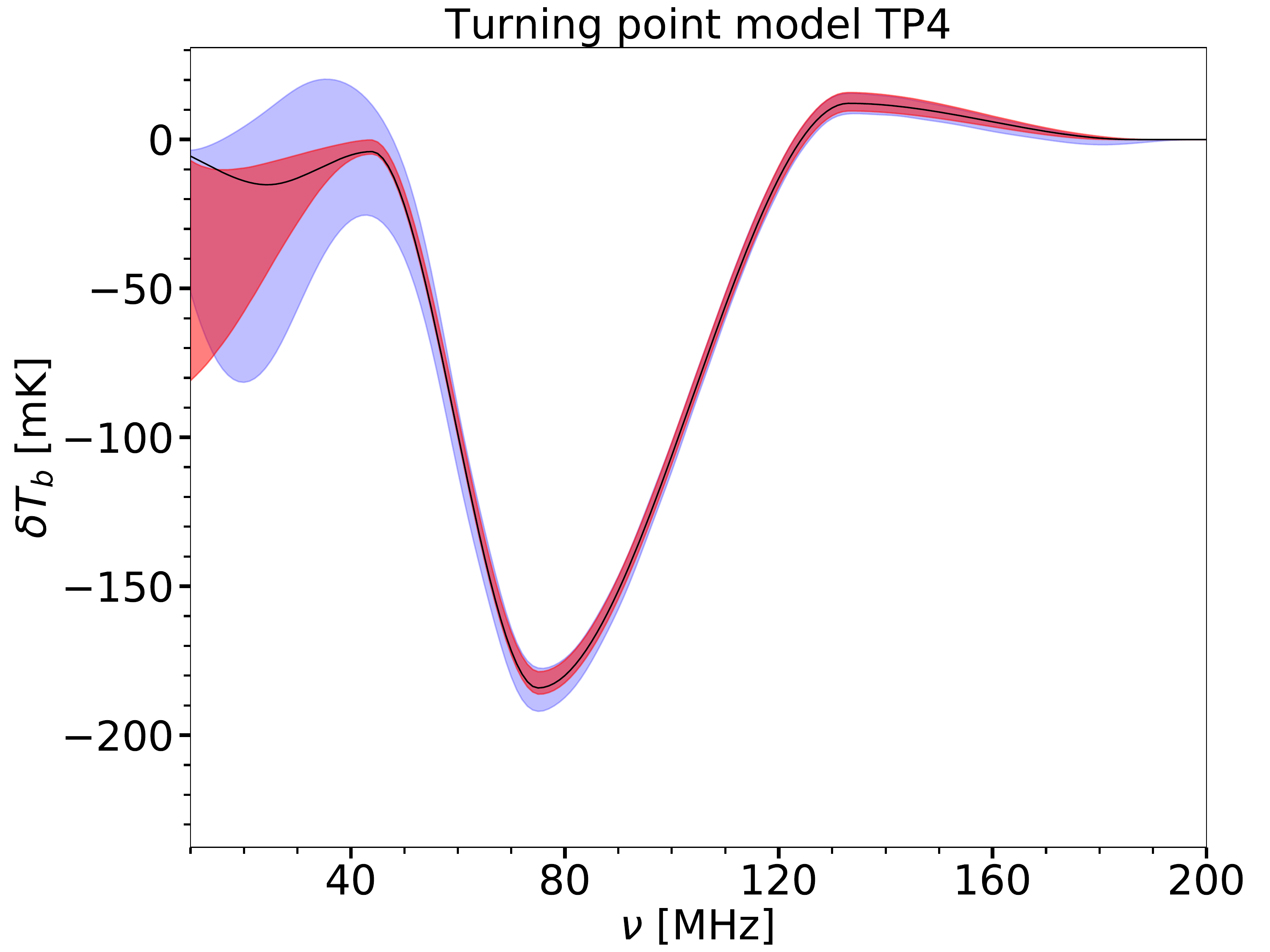}
\caption{Pipeline constraints in frequency space for four random turning point models.~In all panels, the blue intervals correspond to $95\%$ confidence intervals from the linear fit, which uses SVD modes to represent the signal in addition to the foreground, while the red bands correspond to $95\%$ confidence intervals from the MCMC fit, which uses the full nonlinear signal model and SVD foreground modes. For the linear fit, the $95\%$ confidence intervals correspond to $2.5\sigma$ (see Figure~\ref{fig:linear-calibration}).}
\label{fig:turning-point-frequency-space}
\end{center}
\end{figure*}

Figure~\ref{fig:flattened-gaussian-frequency-space} shows the success of the pipeline for the flattened Gaussian model while Figure~\ref{fig:turning-point-frequency-space} shows the same for the turning point model. In each figure, the blue (red) regions indicate linear (MCMC) constraints on the signals in frequency-brightness temperature space.

For the flattened Gaussian (FG) model (Figure~\ref{fig:flattened-gaussian-frequency-space}), the MCMC constraints are clearly tighter than those obtained from the linear fits. However, this difference varies considerably among the cases displayed, being most extreme in case number 3, FG3, where the signal is relatively well localized in frequency, as it is also in cases FG1 and FG5, but closer to the edge of the frequency band than these two cases. On the other hand, signals FG2 and FG4 are wider, particularly FG4 for which the difference in constraining power between the linear and MCMC fits is the smallest, and FG2 is also flatter (i.e. has a larger value of $\tau$) than the rest.

For the turning point (TP) model (Figure~\ref{fig:turning-point-frequency-space}), the four random cases tested also present overall tighter reconstructions for the MCMC fits at the end of the pipeline, but the differences between the MCMC and linear fits are smaller than for most of the flattened Gaussian cases (FG1-3, FG5). All of the turning point model cases and FG4 span wide fractions of the frequency band, leading to larger MCMC uncertainties than the other signals.

\begin{table}[t!]
    \centering
    \caption{Numbers of SVD modes used for the linear fits and their RMS errors}
    \begin{tabularx}{\columnwidth}{@{\extracolsep{\stretch{1}}}*{4}{c}@{}}
        \hline
        \hline
        Model & Signal terms & Foreground terms & RMS (mK) \\
        \hline
        FG1 & 15 & 8 & 3 \\
        FG2 & 34 & 10 & 9 \\
        FG3 & 40 & 11 & 19 \\
        FG4 & 21 & 8 & 9 \\
        FG5 & 17 & 9 & 4 \\
        \hline
        TP1 & 33 & 22 & 7 \\
        TP2 & 30 & 20 & 6 \\
        TP3 & 32 & 24 & 7 \\
        TP4 & 31 & 20 & 6 \\
        \hline
    \end{tabularx}
    \vspace{0.8ex}

    {\raggedright \textbf{Notes.} Flattened Gaussian (FG) fits in Figure~\ref{fig:flattened-gaussian-frequency-space}. Turning point (TP) fits in Figure~\ref{fig:turning-point-frequency-space}. RMS errors computed via Equation~\ref{eq:channel-rms}. \par}

    \label{tab:terms-and-rms}
\end{table}

In comparison with FG, the absorption features in the TP cases cover more similar frequency ranges between each other because they are built based on the thermal history milestones described in Section~\ref{sec:turning-points}.~This physically-motivated similarity between spectral shapes can be seen in the training set sample of the TP model shown in the right panel of Figure~\ref{fig:training-sets}, particularly compared with that of the left panel of this figure for the FG model.

\begin{figure*}[tb]
\begin{center}
\includegraphics[width=0.49\textwidth]{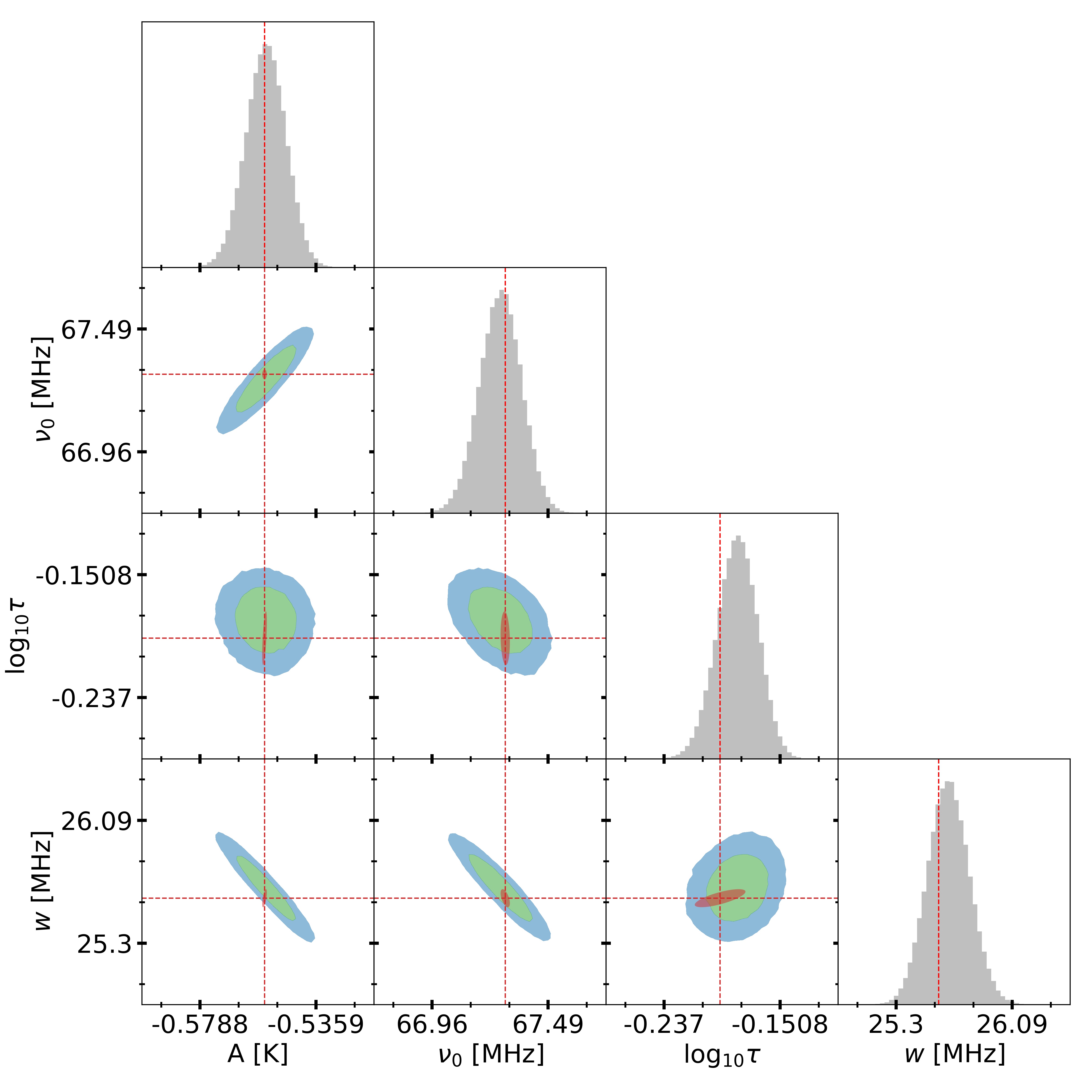}
\includegraphics[width=0.49\textwidth]{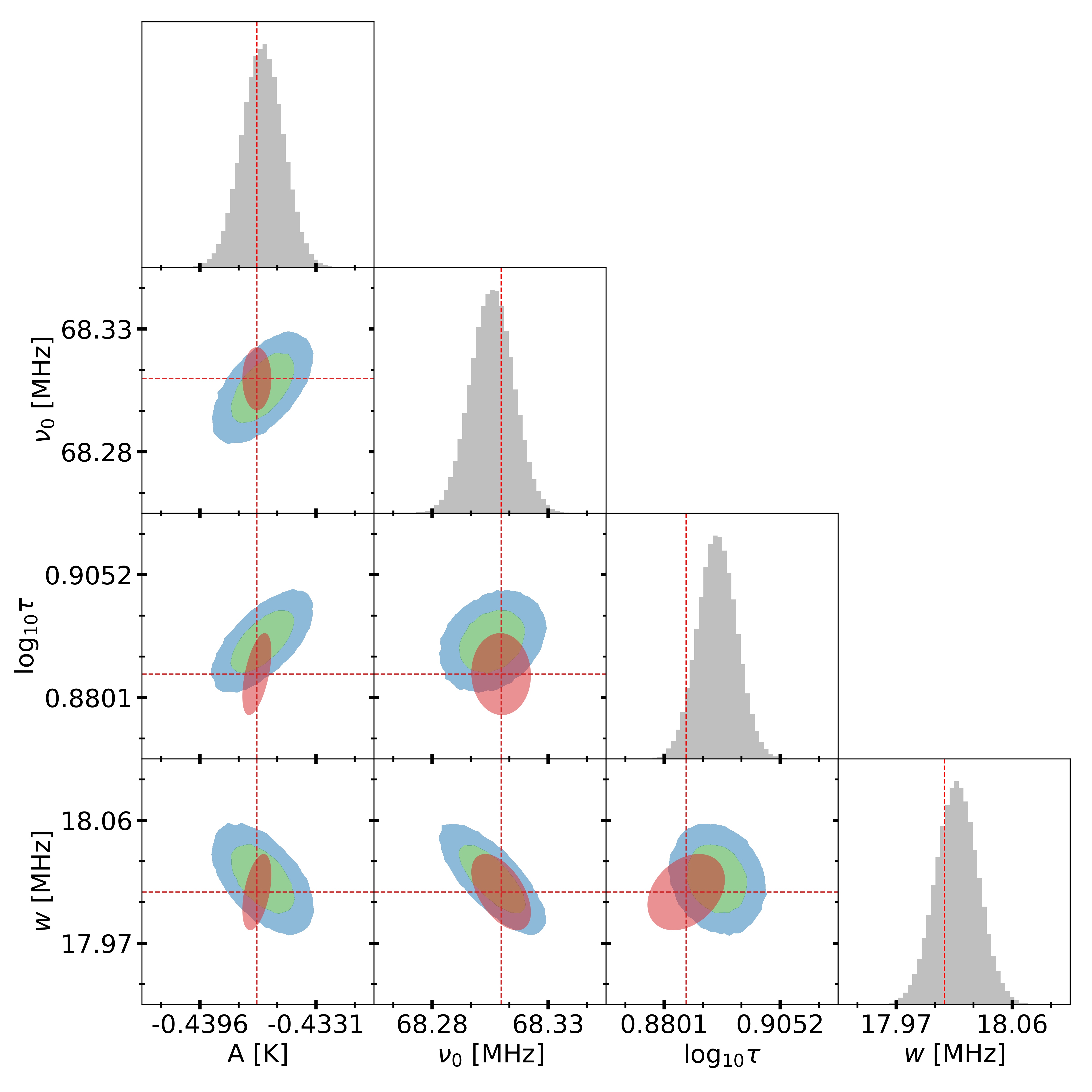}
\caption{1D and 2D MCMC posterior distributions for the flattened Gaussian parameters, with red, dashed lines marking the input parameters. The left (right) triangle plot shows constraints for the signal case FG4 (FG2) in Figure~\ref{fig:flattened-gaussian-frequency-space} and Table~\ref{tab:flattened-gaussian-recovered-parameters}. In the 2D plots, the blue and green contours show 68\% and 95\% confidence intervals, respectively, and the red contours represent the 95\% confidence regions obtained from a Fisher matrix covariance of these parameters for the statistical, radiometer noise. These Fisher matrix estimates assume that systematics are subtracted out perfectly, as if the signal was being observed in isolation with only noise obscuring it.~Clearly, in case FG4 (left), systematics play a more important role in expanding the posterior uncertainties than in FG2 (right).}
\label{fig:flattened-gaussian-triangle-plot}
\end{center}
\end{figure*}

\subsubsection{Numbers of SVD modes}
\label{sec:numbers_modes}

An estimate of the number of SVD modes needed for each training set (foreground and signal) is shown in Figure~\ref{fig:signal-mode-importances} of Appendix~\ref{app:eigenvalue-spectra}.~For the FG signal model, fitting all curves of the signal training set below the noise level requires a large number of SVD terms (see the corresponding calculation in Appendix~\ref{app:eigenvalue-spectra}), of the order of 40 (top, left panel), while for the TP signal model all curves in the training set can be fitted with significantly fewer SVD terms, about 20 (top, right panel).

Given that for each signal model fit, the foreground training set used is the same within the frequency range in common (50-100 MHz), the significant differences in shape among the curves shown in the left panel of Figure~\ref{fig:training-sets} (e.g., central frequency, width, and flattening factor of the troughs), require a relatively large number of terms to describe the entire FG model training set. This will generally imply larger frequency band errors when reconstructing individual signals, in particular for curves with little overlap with the rest of the training set, such as FG3 at the edge of the frequency range in Figure~\ref{fig:flattened-gaussian-frequency-space}.

\begin{table*}[t!]
    \centering
    \caption{MCMC-derived 99\% confidence intervals on the flattened Gaussian parameters} \label{tab:flattened-gaussian-recovered-parameters}
    \begin{tabularx}{\textwidth}{@{\extracolsep{\stretch{1}}}*{6}{c cc cc cc cc cc}@{}}
        \hline
        \hline
        & \multicolumn{2}{c}{FG1} & \multicolumn{2}{c}{FG2} & \multicolumn{2}{c}{FG3} & \multicolumn{2}{c}{FG4} & \multicolumn{2}{c}{FG5} \\
        Par. & Input & Recovered & Input & Recovered & Input & Recovered & Input & Recovered & Input & Recovered \\
        \hline
        $A$ (mK) & -389.3 & $-389.2_{-2.0}^{+2.0}$ & -436.4 & $-436.1_{-3.0}^{+3.0}$ & -551.4 & $-551.7_{-1.2}^{+1.2}$ & -555 & $-555_{-20}^{+19}$ & -755.5 & $-753.9_{-1.9}^{+1.9}$ \\
        \hline
        $\nu_0$ (kHz) & $79914$ & $79911_{-12}^{+12}$ & $68309$ & $68305_{-24}^{+24}$ & $88824.2$ & $88824.1_{-4.5}^{+4.5}$ & $67300$ & $67270_{-250}^{+240}$ & $84953.1$ & $84953.8_{-5.1}^{+5.1}$ \\
        \hline
        $w$ (kHz) & 10295 & $10304_{-38}^{+38}$ & 18006 & $18016_{-44}^{+44}$ & 8570 & $8571_{-12}^{+12}$ & 25590 & $25660_{-360}^{+380}$ & 9383 & $9380_{-17}^{+17}$ \\
        \hline
        $\tau$ & 0.712 & $0.711_{-0.060}^{+0.061}$ & 7.67 & $7.79_{-0.19}^{+0.20}$ & 3.202 & $3.191_{-0.041}^{+0.041}$ & 0.638 & $0.657_{-0.059}^{+0.058}$ & 0.235 & $0.267_{-0.031}^{+0.031}$ \\
        \hline
    \end{tabularx}
    \vspace{0.8ex}

    {\raggedright \textbf{Notes.} All fits done with 800 hours of integration. Spectral constraints of these models are shown in Figure~\ref{fig:flattened-gaussian-frequency-space}. FG4 (FG2) corresponds to the triangle plot in the left (right) panel of Figure~\ref{fig:flattened-gaussian-triangle-plot}. \par}
\end{table*}

Comparatively, the TP model presents less variation and more overlap between curves (right panel of Figure~\ref{fig:training-sets}) and thus requires fewer SVD modes to describe the corresponding training set. The bottom panels of Figure~\ref{fig:signal-mode-importances} show the numbers of terms (7 and 12) required to fit the foreground training set curves corresponding to each signal model. In this case, the larger frequency range covered in the TP model (10-200 MHz) is bound to increase the number of SVD terms needed to fit all curves below the noise level.

Note, however, that the number of parameters shown in Figure~\ref{fig:signal-mode-importances} is calculated for the overall training sets. For the individual cases displayed in Figures~\ref{fig:flattened-gaussian-frequency-space} and~\ref{fig:turning-point-frequency-space} for each model, Table~\ref{tab:terms-and-rms} shows the number of terms chosen by the DIC and the corresponding RMS uncertainties when simultaneously fitting signal and foreground. The individual cases follow the same pattern as is seen in Figure~\ref{fig:signal-mode-importances}, where TP models require more terms for foreground and signal than FG models.

\begin{figure*}[tb]
\begin{center}
\includegraphics[width=0.93\textwidth]{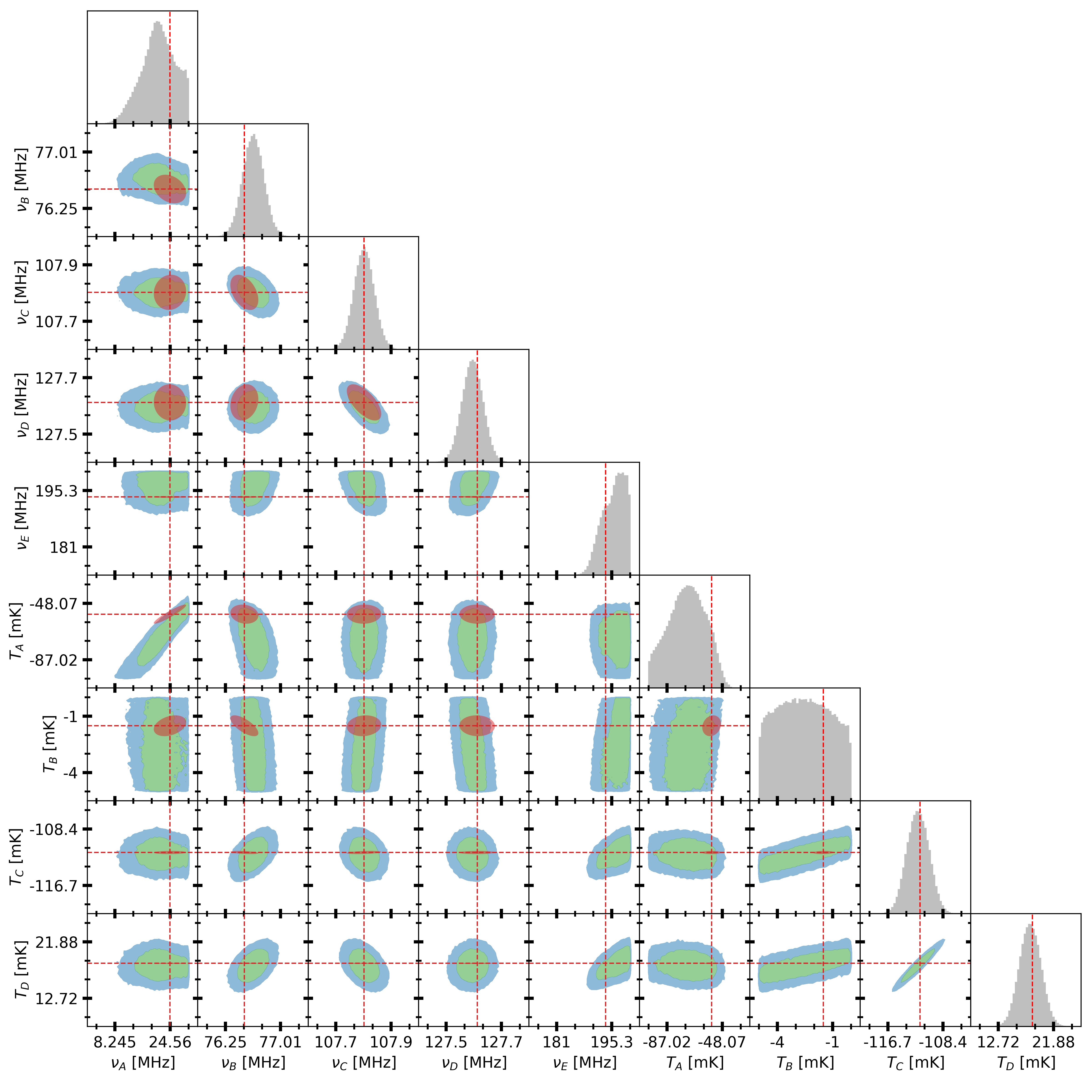}
\caption{Same as Figure~\ref{fig:flattened-gaussian-triangle-plot} but for the turning point model shown in the upper left panel of Figure~\ref{fig:turning-point-frequency-space} (model TP1). Blue and green contours in the 2D plots show 68\% and 95\% confidence intervals, while the red ellipses show Fisher-matrix estimated 95\% confidence intervals which assume only statistical noise. All intervals are for 800 hours of integration. Some parameters, such as the temperature of turning point B, which is only allowed to vary from -5 to 0 mK, are not constrained within the prior volume while others, such as the temperature of turning point C are constrained.}
\label{fig:turning-point-triangle-plot}
\end{center}
\end{figure*}

\begin{table*}[t!]
    \centering
    \caption{MCMC-derived 99\% confidence intervals on the turning point parameters}  \label{tab:turning-point-recovered-parameters}
    \begin{tabularx}{\textwidth}{@{\extracolsep{\stretch{1}}}*{5}{c cc cc cc cc}@{}}
        \hline
        \hline
        & \multicolumn{2}{c}{TP1} & \multicolumn{2}{c}{TP2} & \multicolumn{2}{c}{TP3} & \multicolumn{2}{c}{TP4} \\
        \hline
        Parameter & Input & Recovered & Input & Recovered & Input & Recovered & Input & Recovered \\
        \hline
        $\nu_A$ (MHz) & 24 & $21_{-13}^{+9}$ & 12 & $13_{-11}^{+17}$ & 23 & $18_{-15}^{+9}$ & 24 & $9_{-8}^{+16}$ \\
        \hline
        $T_A$ (mK) & -56 & $-74_{-26}^{+27}$ & -52 & $-46_{-52}^{+35}$ & -34 & $-62_{-37}^{+41}$ & -15 & $-41_{-56}^{+30}$ \\
        \hline
        $\nu_B$ (MHz) & 76.51 & $76.63_{-0.38}^{+0.37}$ & 50.07 & $49.82_{-0.92}^{+0.95}$ & 42.49 & $42.51_{-0.57}^{+0.56}$ & 43.91 & $43.91_{-0.57}^{+0.57}$ \\
        \hline
        $T_B$ (mK) & -1.5 & $-2.5_{-2.4}^{+2.5}$ & -0.5 & $-1.9_{-3.0}^{+1.9}$ & -0.4 & $-2.1_{-2.9}^{+2.1}$ & -4.0 & $-2.4_{-2.6}^{+2.4}$ \\
        \hline
        $\nu_C$ (MHz) & 107.819 & $107.819_{-0.085}^{+0.085}$ & 106.638 & $106.620_{-0.071}^{+0.071}$ & 61.83 & $62.12_{-0.32}^{+0.32}$ & 74.99 & $75.17_{-0.25}^{+0.25}$ \\
        \hline
        $T_C$ (mK) & -111.9 & $-112.1_{-4.1}^{+4.2}$ & -231.8 & $-233.8_{-4.8}^{+5.9}$ & -159.2 & $-156.8_{-6.4}^{+6.2}$ & -184.1 & $-182.5_{-4.8}^{+5.1}$ \\
        \hline
        $\nu_D$ (MHz) & 127.625 & $127.612_{-0.077}^{+0.075}$ & 131.934 & $131.693_{-0.054}^{+0.054}$ & 128.56 & $128.42_{-0.43}^{+0.43}$ & 133.02 & $133.00_{-0.22}^{+0.21}$ \\
        \hline
        $T_D$ (mK) & 18.4 & $18.0_{-4.6}^{+4.5}$ & 6.3 & $4.6_{-4.2}^{+5.6}$ & 14.2 & $14.6_{-1.2}^{+1.3}$ & 12.2 & $12.5_{-3.8}^{+4.9}$ \\
        \hline
        $\nu_E$ (MHz) & 193.63 & $195.69_{-7.9}^{+4.3}$ & 197 & $190_{-25}^{+10}$ & 153.980 & $153.937_{-0.48}^{+0.48}$ & 186 & $185_{-3}^{+12}$ \\
        \hline
    \end{tabularx}
    \vspace{0.8ex}

    {\raggedright \textbf{Notes.} All fits performed with 800 hours of integration. Parameter constraints subject to the priors given in Table~\ref{tab:turning-point-training-set-distribution}, except for that on $\nu_A$ which was allowed to uniformly vary from (1, 30) for extra variability in the MCMC search.~Spectral~constraints on these signal cases are shown in Figure~\ref{fig:turning-point-frequency-space}. TP1 corresponds to the triangle plot in Figure~\ref{fig:turning-point-triangle-plot}. \par}
\end{table*}

\begin{figure*}[tb]
  \centering
  \includegraphics[width=0.43\textwidth]{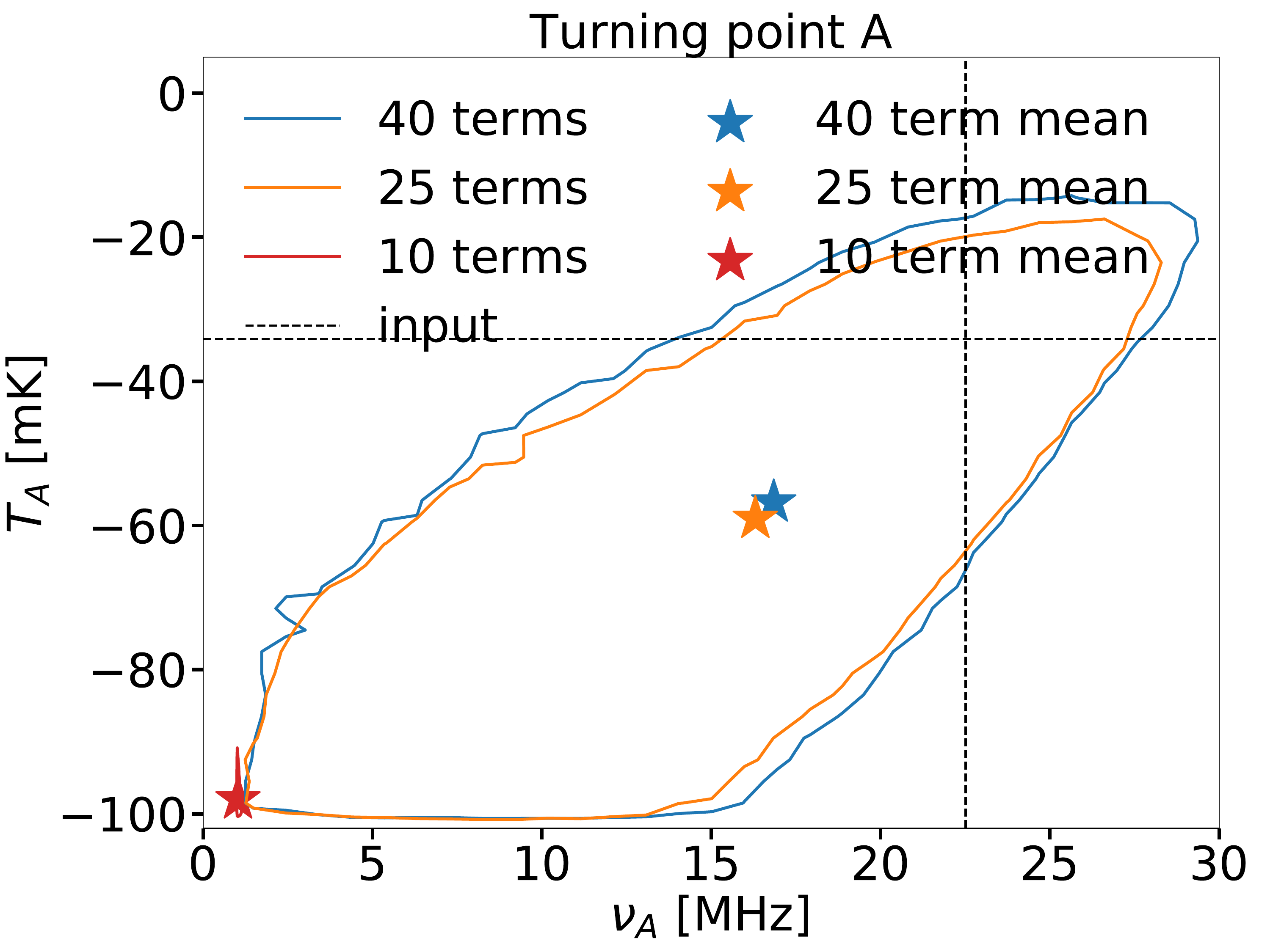}
  \includegraphics[width=0.43\textwidth]{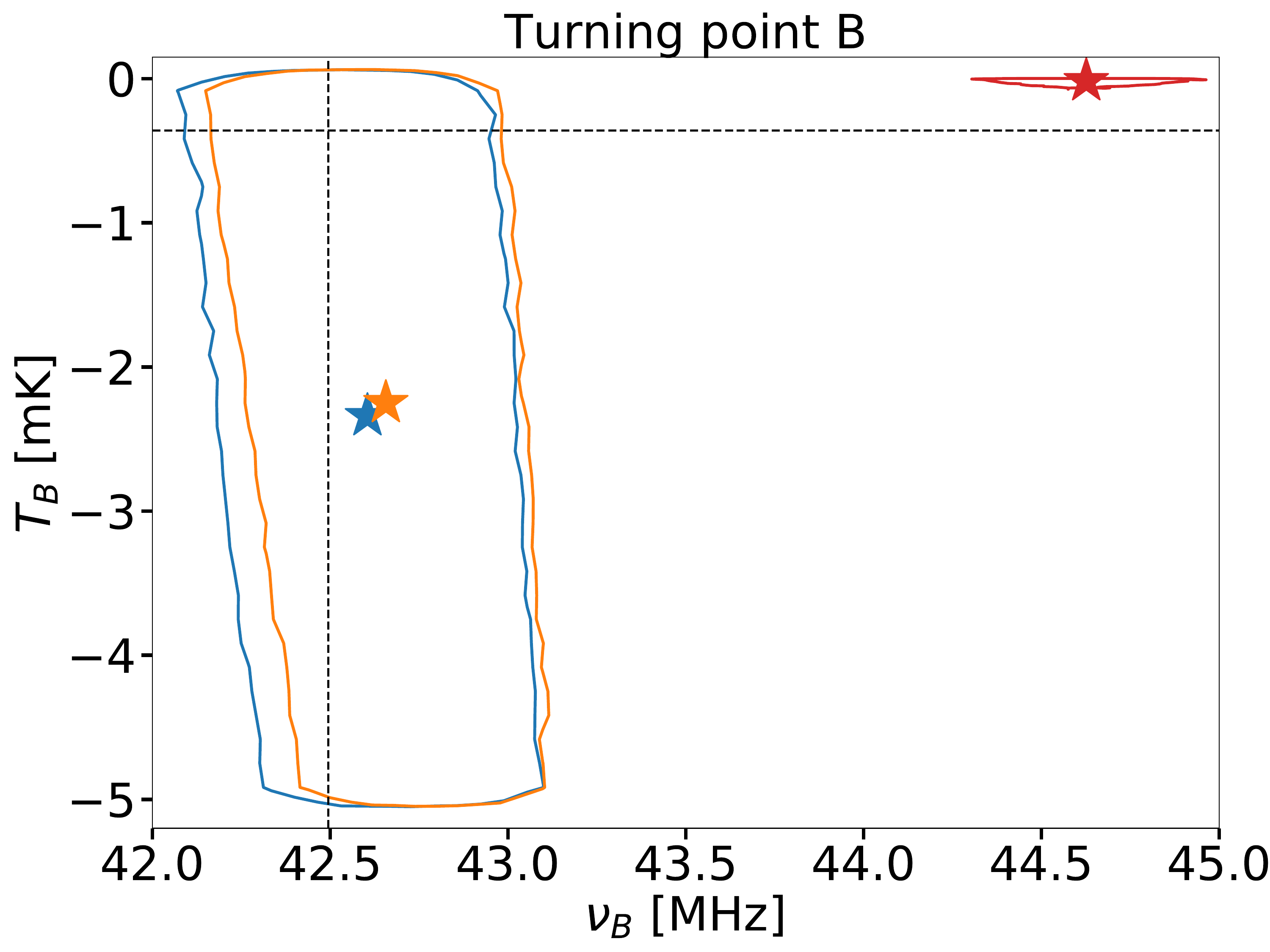}
  \includegraphics[width=0.43\textwidth]{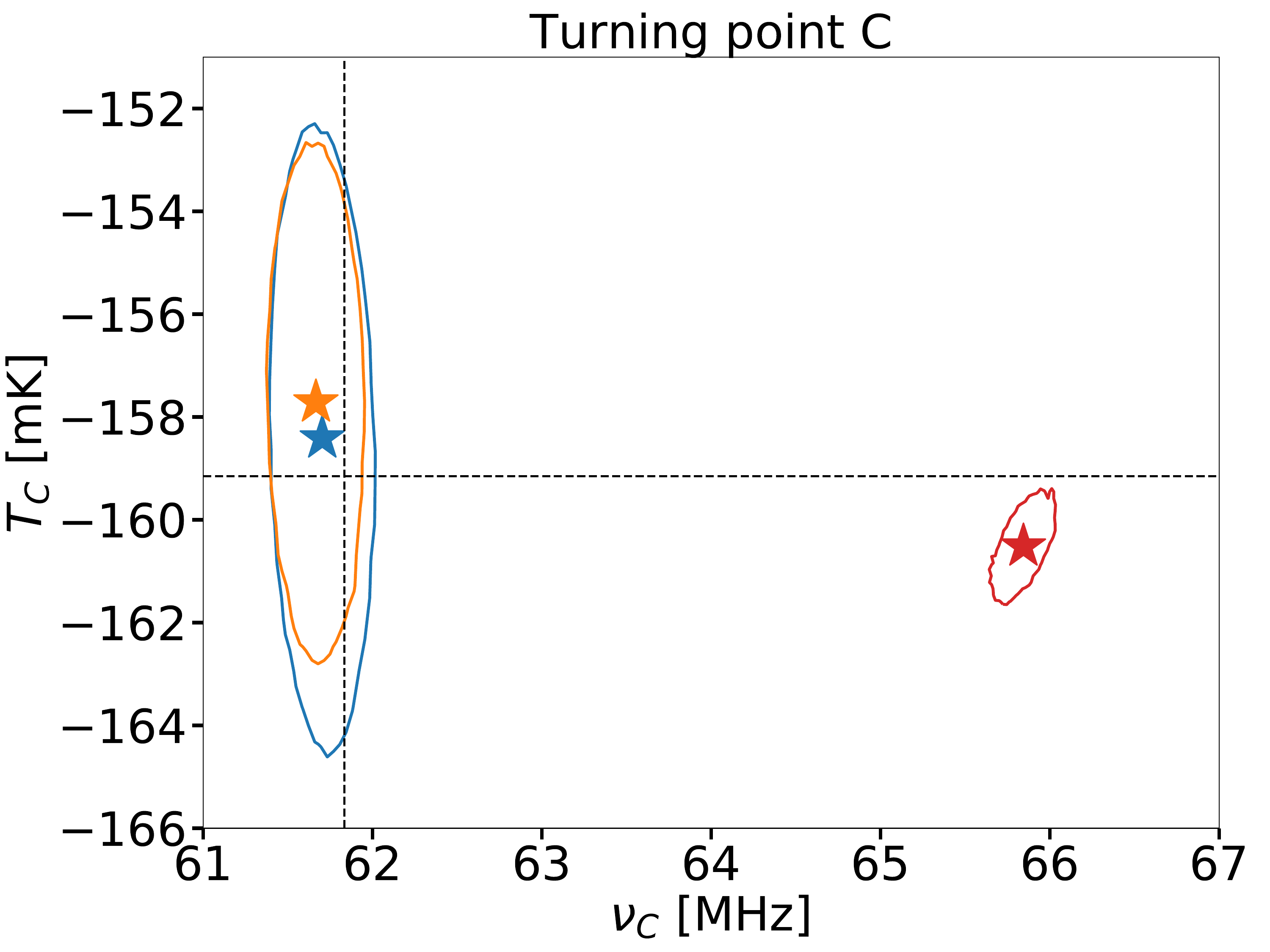}
  \includegraphics[width=0.43\textwidth]{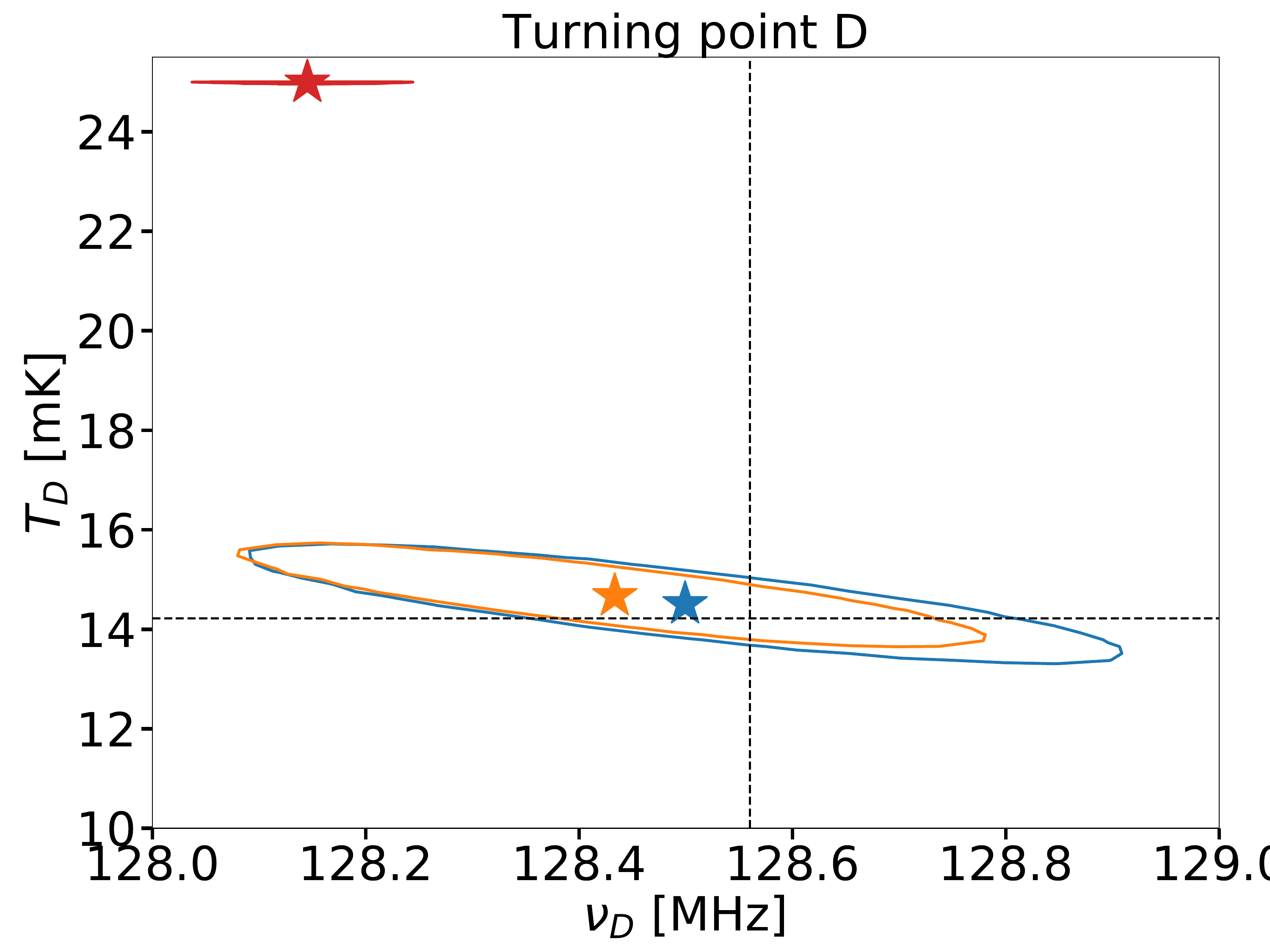}
  \caption{For case TP3, means (stars) and $95\%$ confidence contours for turning points A-D when marginalizing over 10 (red), 25 (orange), and 40 (blue) foreground terms in the MCMC fit.~The dashed vertical and horizontal lines indicate the input parameters. When 10 terms are marginalized over (note that 24 terms were chosen in the linear fit; see Table~\ref{tab:terms-and-rms}), the foreground model is insufficient to explain the data. The resulting constraints (red) are thus biased and spuriously tight (exacerbating the bias) since fewer terms lead to smaller covariances between foreground and signal. For turning points A, B and D, the number of terms is low enough for the red contours to even find edges of the prior volume. When enough terms are included, such as for example 25, the constraints (orange) become unbiased and the uncertainties sufficiently accurate. Adding more terms, such up to 40, the results (blue) asymptote to the target posterior without qualitatively changing thanks to the use of foreground priors described in Section~\ref{sec:foregroundterms}~and~Appendix~\ref{app:priors-from-svd}.}
  \label{fig:constraints-vs-num-terms}
\end{figure*}

\subsubsection{Recovering input parameters}
\label{sec:mcmcconstraints}

The MCMC recovered one dimensional (1D) posterior probability distributions for the flattened Gaussian and turning point model parameters are shown in gray in the diagonal plots of Figures~\ref{fig:flattened-gaussian-triangle-plot}~and~\ref{fig:turning-point-triangle-plot} respectively, where the red, dashed lines mark the input parameters. The contours in the two dimensional (2D) off-diagonal plots in these figures show the $68\%$ (green) and $95\%$ (blue) confidence levels and present the covariances between these parameters as found by the MCMC calculation. In red contours, for comparison, we show $95\%$ confidence level contours for Fisher-matrix-derived covariances (see Section~\ref{sec:fisher_matrix}) from the statistical noise of the radiometer equation.

We find that our pipeline successfully obtains constraints consistent with the input values for both models within the noise level (red contours), and it does so efficiently by rapidly reaching both the targeted acceptance rate (25\%) for the MCMC sampler (see Section~\ref{sec:updating} and Appendix~\ref{app:updating}) and ultimately a high level of convergence for all parameters, as determined by the commonly employed Gelman-Rubin test~\citep{Gelman:92}.

Figure~\ref{fig:flattened-gaussian-triangle-plot} shows constraints for two cases, FG4 (left panel) and FG2 (right) from Figure~\ref{fig:flattened-gaussian-frequency-space}, which are representative of two types of results. For case FG2 (right), the 95\% confidene level MCMC constraints (blue contours) are much tighter than those obtained with the linear fit (see Figure~\ref{fig:flattened-gaussian-frequency-space}), reaching down almost to the noise level (red contours). On the other hand, the MCMC constraints (blue) for case FG4 (left) are noticeably larger than those for FG2. Note also that the non-linear constraints of FG4 (blue) are significantly larger than the corresponding Fisher matrix contours (red) derived only from the 800 hours of integration, i.e. the noise level. This indicates that for FG4 the overlap between signal and foreground is larger than for FG2.

The difference between the actual constraints in blue contours, and those in red for the noise level reference, corresponds to the effect of simultaneously fitting the signal together with the foreground. This is necessary to fit the data consistently, properly accounting for systematic uncertainties on the signal caused by the combination of random noise and large foreground systematics. Thus, by comparing cases FG4 and FG2, it is clear that FG4 suffers from larger systematic errors due to its larger overlap with the foreground training set.

Figure~\ref{fig:turning-point-triangle-plot} shows the constraints on the turning point model parameters for case TP1. Note that there are large differences between systematic (blue contours at the 95$\%$ confidence level) and statistical (red) errors among some of the parameters and that some parameters (e.g. $\nu_A$, $\nu_E$, and $T_B$) hit edges of their prior space.

As a summary of all MCMC fits performed for the FG and TP cases presented in Figures~\ref{fig:flattened-gaussian-frequency-space} and~\ref{fig:turning-point-frequency-space}, Tables~\ref{tab:flattened-gaussian-recovered-parameters}~and~\ref{tab:turning-point-recovered-parameters} show 99\% confidence intervals on their respective parameters.

\begin{figure*}[tb]
  \centering
  \includegraphics[width=0.5\textwidth]{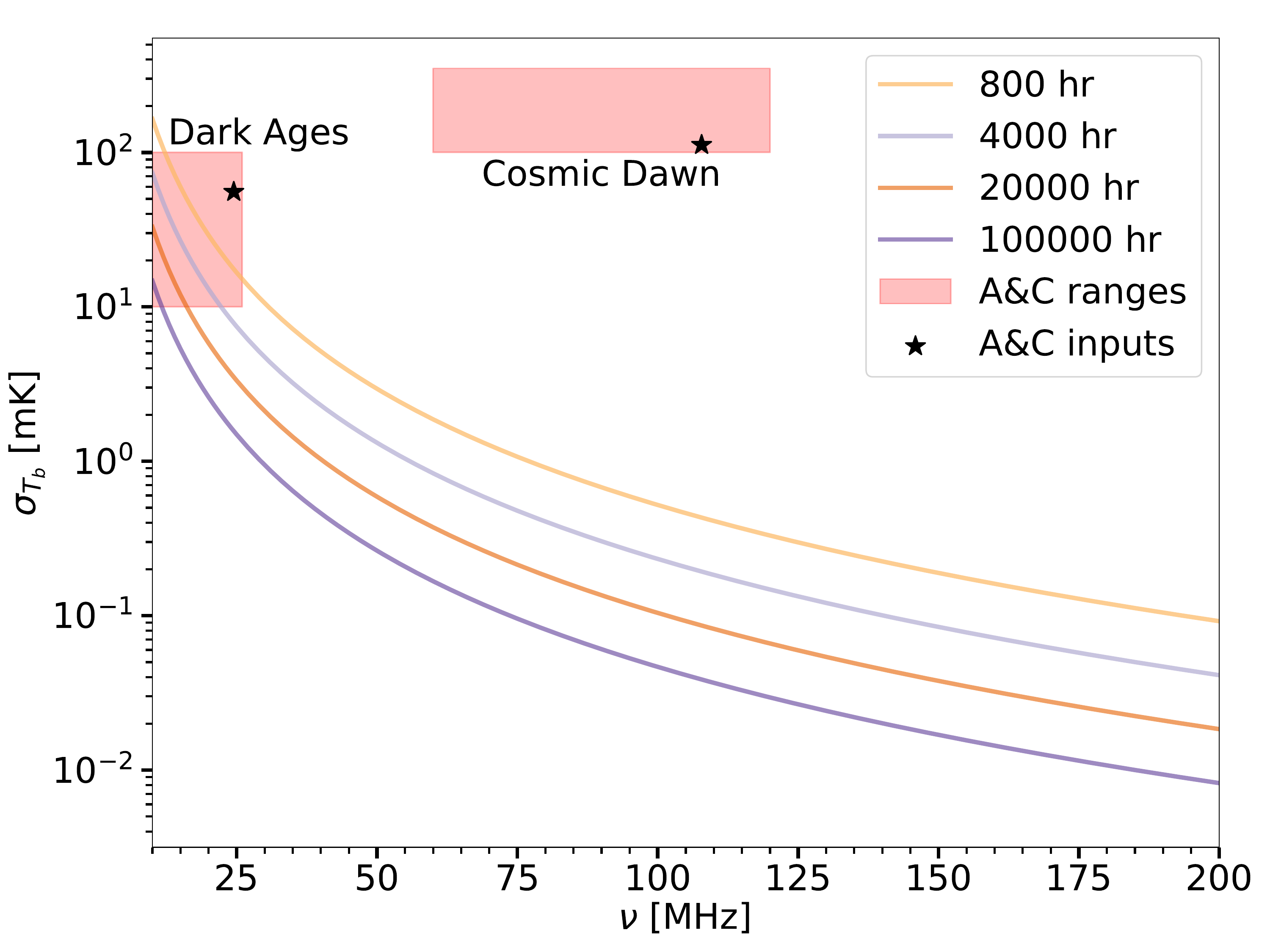}
  \includegraphics[width=0.49\textwidth]{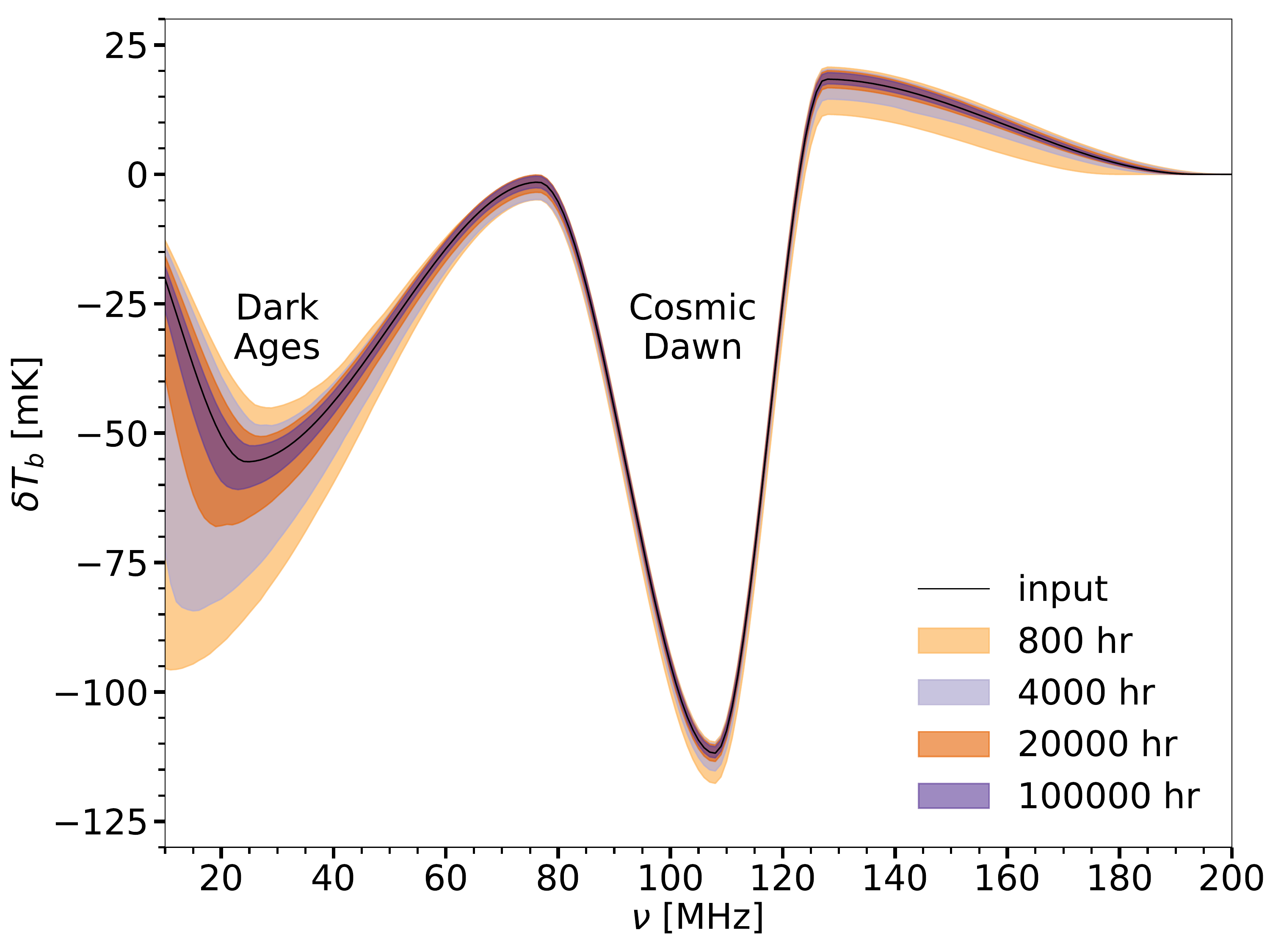}
  \caption{\textit{Left}: $1\sigma$ noise levels for the integration times used in the application example of Section~\ref{sec:integration-time}.~The red rectangles encompass the allowed ranges of frequencies and absolute values of temperatures for turning points A and C, representing the Dark Ages and Cosmic Dawn troughs respectively. The black stars indicate the input values of the frequencies and temperatures of these turning points for signal TP1.~The uncertainties given by the curves correspond only to the noise levels and do not account for uncertainties deriving from the overlap between foreground and signal.~\textit{Right}: Full (statistical plus systematic) uncertainties of TP1 in frequency space for four evenly increasing integration times up from our reference value of 800 hours (see the top, left panel of Figure~\ref{fig:turning-point-frequency-space}), with the same random seed for noise generation to ensure comparability.} \label{fig:noise-level-vs-integration-time}
\end{figure*}

\begin{figure}[tb]
  \centering
  \includegraphics[width=0.23\textwidth]{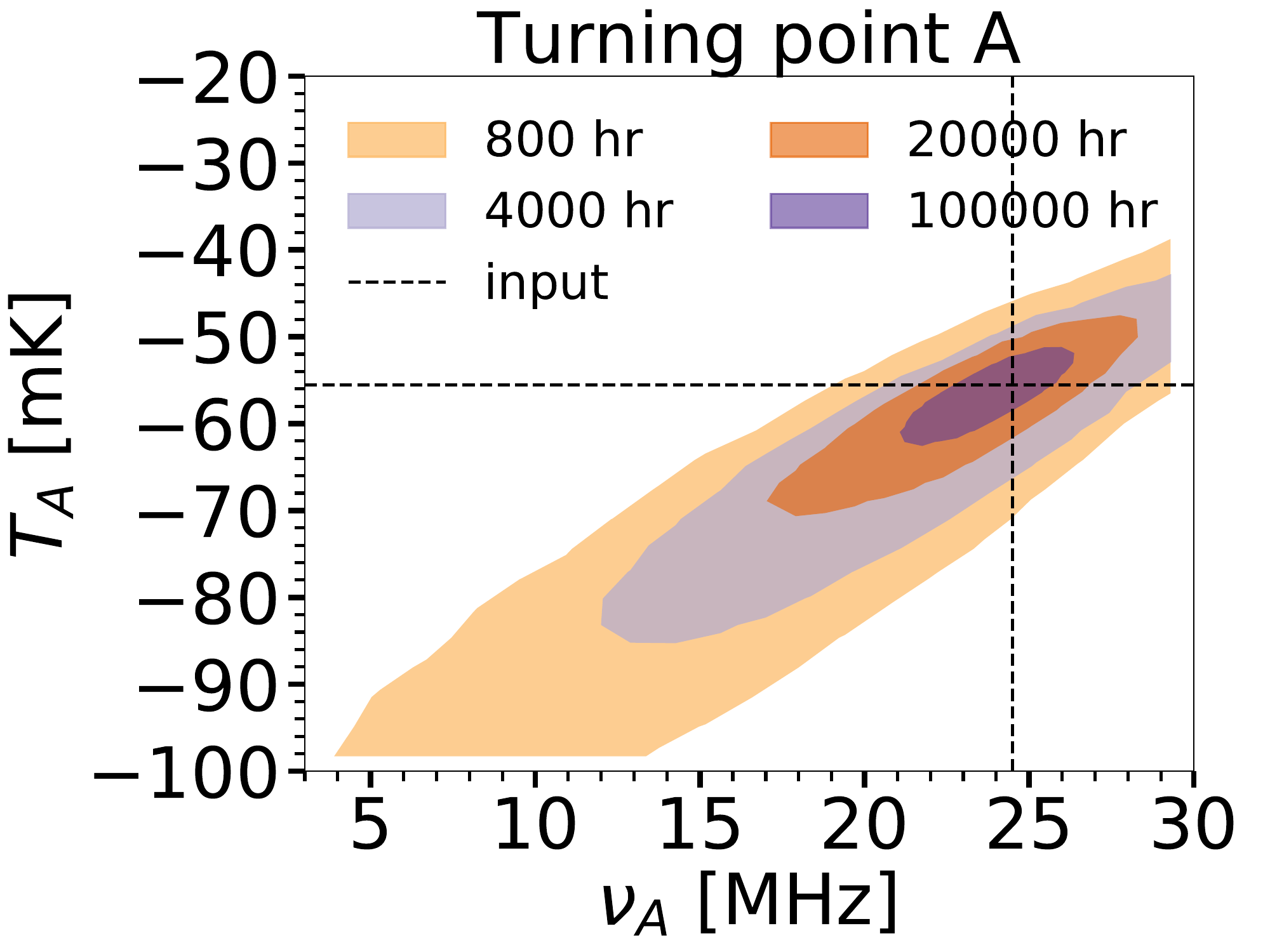}
  \includegraphics[width=0.23\textwidth]{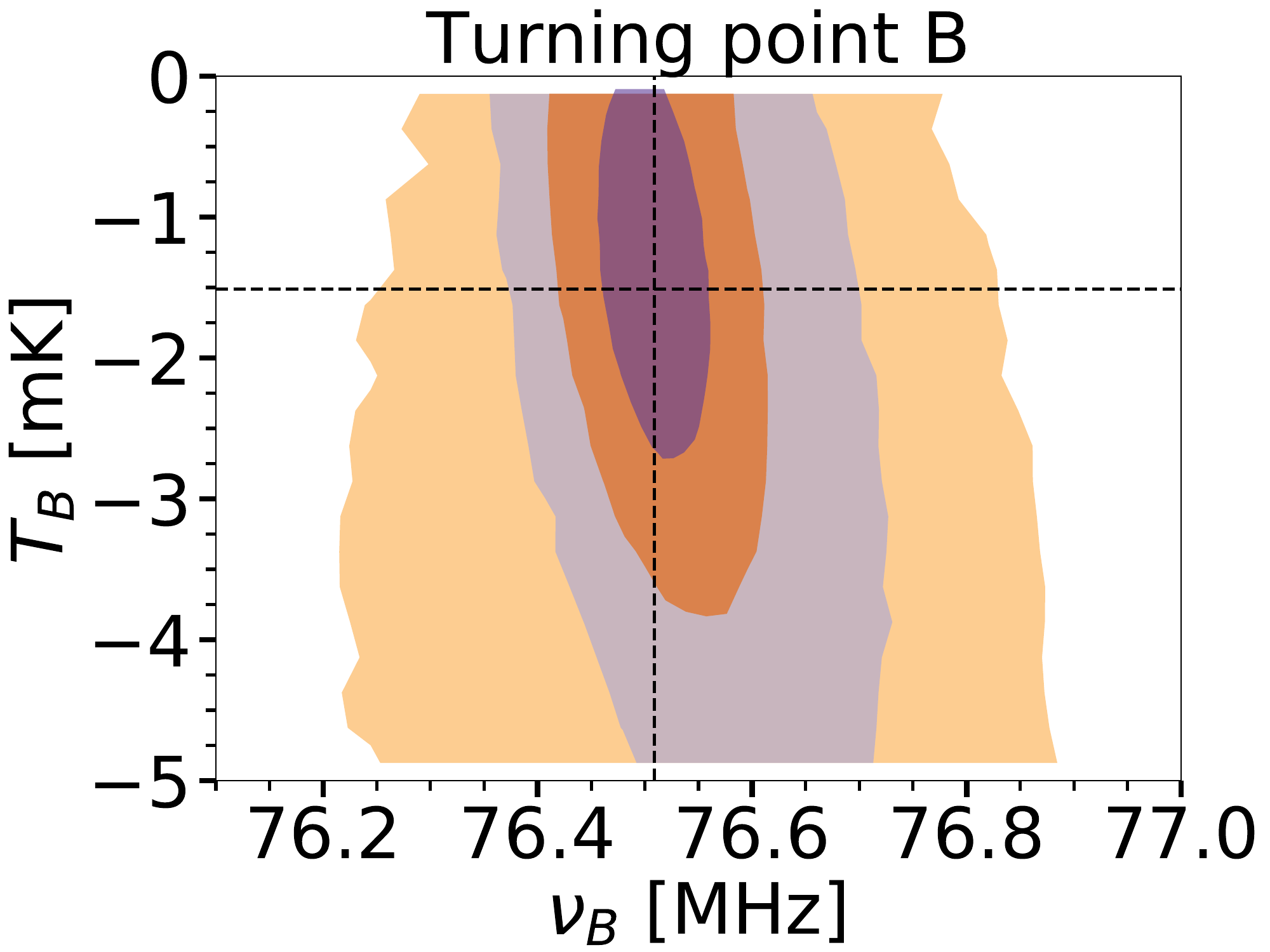}
  \includegraphics[width=0.23\textwidth]{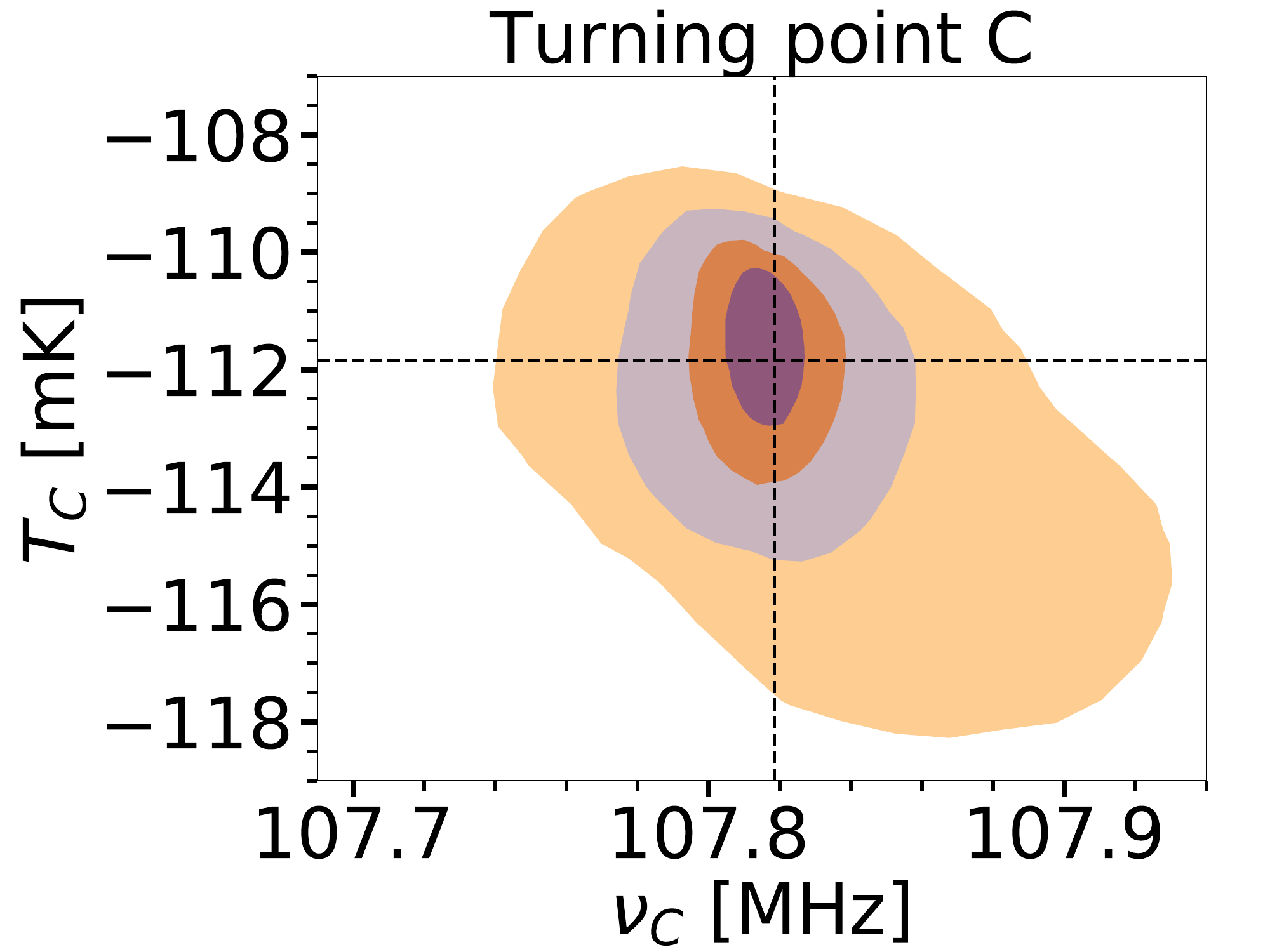}
  \includegraphics[width=0.23\textwidth]{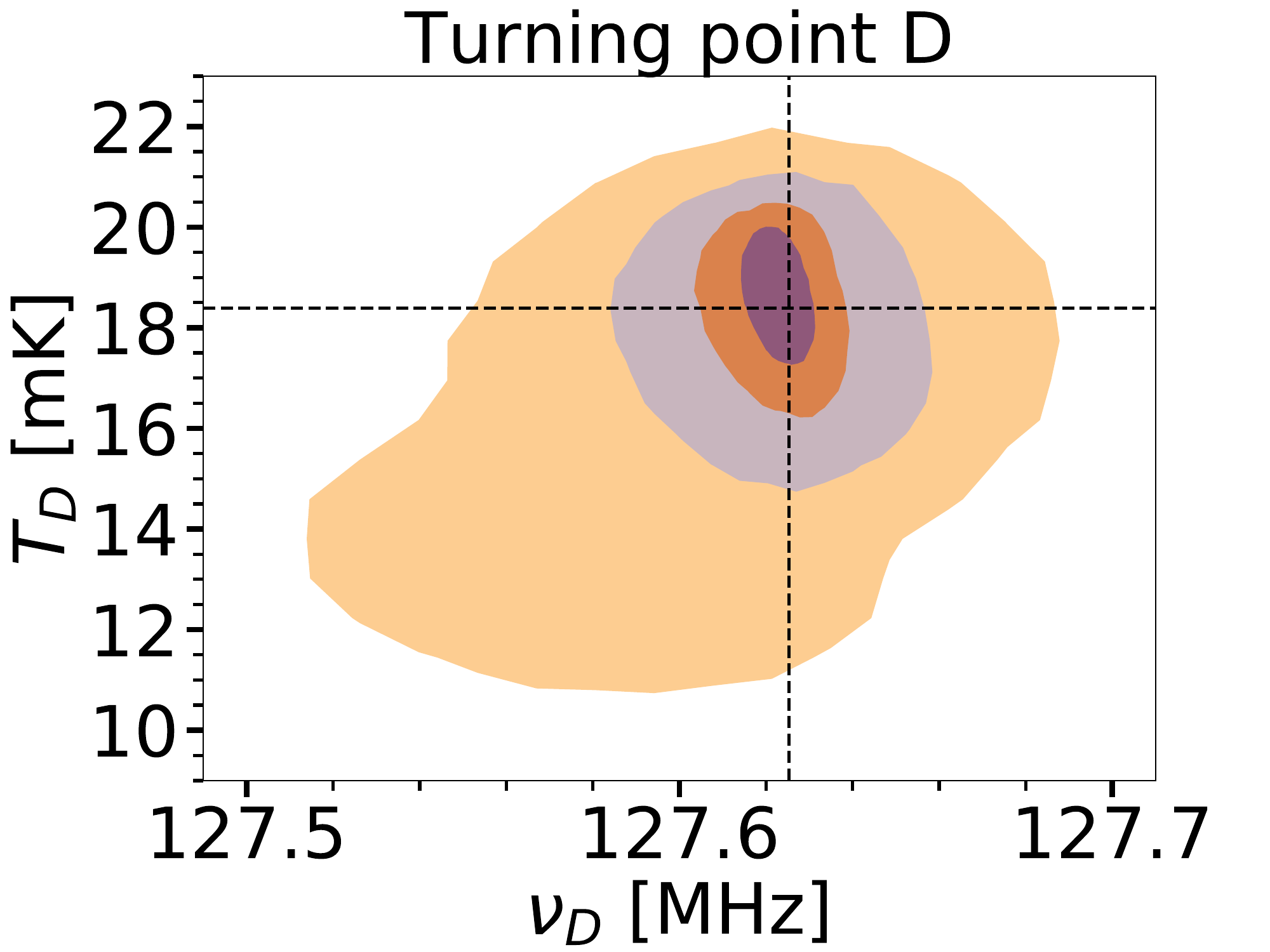}
  \caption{$95\%$ confidence level constraints on the frequencies and temperatures of turning points A-D versus integration time.~The same random seed is used for each case, meaning that the noise has the same shape but a different magnitude, given by the integration time. The input signal for these fits is TP1, as in Figure~\ref{fig:turning-point-triangle-plot}.~The colors for the integration times match those of Figure~\ref{fig:noise-level-vs-integration-time}.} \label{fig:constraints-vs-integration-time}
\end{figure}

\subsection{Number of marginalized foreground parameters}
\label{sec:varying-num-terms}

As an important test to be passed by our new methodology, in Figure~\ref{fig:constraints-vs-num-terms} we show the change in constraints when we vary the number of SVD foreground terms marginalized over in the MCMC analysis of turning point model 3 (TP3). As described in Section~\ref{sec:refmodel}, Table~\ref{tab:terms-and-rms} indicates that this specific realization of TP3 requires 24 foreground terms.~Based on this reference value, we calculate constraints on the TP model parameters when using three different numbers of foreground terms: 10, 25 and 40. Given the reference value, we expect to have highly biased means and spurious uncertainties for the 10 terms case, in contrast with those of the 25 and 40 term cases. This is actually shown in Figure~\ref{fig:constraints-vs-num-terms} for the $95\%$ confidence level constraints on A, B, C and D in frequency-temperature space. In addition, the uncertainties of the 25 and 40 term cases are remarkably similar between each other despite the large difference in number of parameters. This indicates the success of our technique of employing training set priors (see Section~\ref{sec:foregroundterms} and Appendix~\ref{app:priors-from-svd}) to avoid the contributions of SVD foreground modes with SVD importances below the noise level. Thus, our technique only requires selecting a number of foreground terms large enough above that found by the linear fit to provide unbiased, accurate and robust parameter measurements.

\subsection{Constraints vs. integration time}
\label{sec:integration-time}

In this section, we utilize our new pipeline to run fits for turning point model TP1 when evenly increasing the integration time by factors of 5 from 800 (see the top, left panel of Figure~\ref{fig:turning-point-frequency-space}) to 4000, 20000, and 100000 hours.~These times provide the spectral noise profiles shown in the left panel of Figure~\ref{fig:noise-level-vs-integration-time}, while the right panel shows the corresponding increases in constraining power on the spectral shape of the signal. These results show that up to the highest integration time used the constraints are not limited by systematics, which in this case is the overlap of the signal with the foreground.

The 2D constraints at 95$\%$ confidence level on each pair of frequency and temperature parameters for turning points A, B, C and D are shown in Figure~\ref{fig:constraints-vs-integration-time}. Consistently with the right panel of Figure~\ref{fig:noise-level-vs-integration-time}, these constraints are increasingly tighter with longer integration times and, as discussed above, no systematic floor is found up to $10^5$ hours.~Note that for each of the four runs the same noise seed was used to have an identical noise shape with the magnitude scaled as $1/\sqrt{t}$. This ensures that the constraints for these four runs are exactly comparable in terms of mean and covariance shape and only the size changes as a function of integration time.

This exercise exemplifies a straightforward application of our pipeline to simulate experimental setups. Given a training set for each of the data components, the pipeline can establish whether a certain amount of integration time reaches or not the systematic floor for the modeling used. In our idealized example, we learn that we could significantly tighten constraints on the Dark Ages trough, with a foreground level considerably higher than Cosmic Dawn, by for instance adding single dual antennas (assuming that they are similar enough to allow comparable calibrations) to efficiently increase the integration time. In upcoming studies, including additional data components such as an instrument (Paper IV) will continue this line of research and provide further applications for the utilization of our pipeline.

\section{Conclusions}
\label{sec:conclusions}

This is the second paper in a series presenting a complete analysis pipeline for global 21-cm observations that accounts for each of the data components simultaneously and self-consistently. In this work, we have advanced the analysis by incorporating the ability to not only separate the signal from foregrounds, as described in the first paper, but also start a conditional MCMC exploration of signal model parameters based on the mean and covariance of the spectral signal constraint derived analytically in the first step.

In each step of the MCMC search over the signal parameter space, we utilize the linear, optimal description of the foreground obtained in the initial SVD calculation to marginalize over its SVD modes analytically, instead of exploring them numerically with the MCMC. As to be expected, this greatly improves the efficiency of the MCMC.~We also implement the use of priors derived from the foreground training set to ensure that the variations which are deemed unimportant by the SVD analysis do not unduly affect our results. This allows us to select a number of SVD foreground modes that is large enough to avoid biases without unnecessarily increasing the uncertainties.

We demonstrate that this technique successfully recovers input parameters for two analytical models. The first is inspired by the recent results from the EDGES collaboration, where a flattened Gaussian shape was employed to fit the observations, together with various foreground models~\citep{Bowman:18}. The other signal model builds upon a well known simplified theoretical description of the global 21-cm spectral form based on predicted extrema (turning points) caused by cosmological and astrophysical phenomena~\citep{Pritchard:10} during the end of the Dark Ages, Cosmic Dawn, and the Epoch of Reionization. For the purpose of testing the pipeline, both of these models are allowed to vary beyond the adiabatic cooling limit of the standard model, as suggested by the EDGES results.

Using a particular case for the turning point model, we test that when varying the number of foreground parameters with respect to the reference value selected by the DIC in the linear fit our technique does behave as intended. That is, when not using a sufficient number of terms we predictably obtain large biases and spuriously tight constraints, and when correctly using a number close to or larger than that from the DIC we are able to reproduce the input values with uncertainties that properly include the statistical noise plus a systematic error accounting for the overlap between the signal and foreground modeling.~The latter varies for individual cases and training sets and is crucially captured by our self-consistent analysis.

It is also worth noting that if choosing a number of terms much larger than that from the DIC the errors do not undesirably increase thanks to our use of priors derived from the foreground training set.~These incorporate the knowledge on the importance of each mode of variation, providing automatic downweighting of irrelevant modes.

As an initial application, we then employ our newly verified pipeline to examine how increasing the integration time affects the constraints on a turning point model case.~Our framework allows us to straightforwardly model the foreground and beam realistically by building informed training sets, as to be presented in currently ongoing work, as well as, in the same manner, to incorporate a full receiver model (see upcoming Paper IV of this series for such an analysis). However, for the idealized foreground and beam used here as test examples, we find no systematic floor on constraining the Dark Ages and Cosmic Dawn troughs when increasing up to a factor of 125 our reference, modest time of 800 hours. This simple exercise serves as an instance of the opportunities opening to the hydrogen cosmology community in utilizing our publicly available, statistically robust pipeline to rigorously analyze both simulations, in preparation of experiments, and actual observations. 

Critically, our analysis technique fully accounts for covariances and systematic uncertainties as encoded in detailed, readily changeable training sets for each of the components forming a given set of sky-averaged 21-cm measurements.~This also implies that no analytical modeling is required.~In its absence, additional measurements and/or simulations can be employed to construct training sets.

Due to the fact that foreground parameters are analytically marginalized instead of being numerically sampled by the MCMC engine, a large number of foreground parameters can be added without loss of efficiency, allowing for many unaveraged spectra to be processed by the pipeline with negligible added computational complexity. Furthermore, it is useful to note that our methodology can be directly adapted to any type of data, and is especially beneficial when covariances between signal and systematics are relatively large.

Planned observations with EDGES, CTP, SARAS, LEDA, PRIzM, REACH and other experiments from the ground, as well as with DAPPER from lunar orbit, should greatly benefit from the framework described here, which pioneeringly combines linear pattern recognition with nonlinear Bayesian statistics.

\acknowledgments{We thank Eric Switzer for useful discussions at the beginning of this project, and for the seed idea of marginalizing over foreground parameters.~D.R. was supported by a NASA Postdoctoral Program Senior Fellowship at the NASA Ames Research Center, administered by the Universities Space Research Association under contract with NASA.~The theoretical work was partially funded by NASA ATP grant NNX15AK80G. J.M. acknowledges support from a CITA National Fellowship. This work is directly supported by the NASA Solar System Exploration Virtual Institute cooperative agreement 80ARC017M0006. Resources supporting this work were provided by the NASA High-End Computing (HEC) Program through the NASA Advanced Supercomputing (NAS) Division at Ames Research Center.}

\appendix

\section{Calibrating confidence intervals for linear fits} \label{app:linear-calibration}

As described in Paper I, the confidence intervals from our linear fits do not match the common 68-95-99.7 percent rule for 1, 2, and 3 sigma implied by the $\chi$ distribution. Instead, these confidence intervals depend on the training sets and the selected numbers of terms. In a nutshell, we perform fits with many foreground, signal, and noise realizations. For each fit, we calculate the signal bias statistic defined as
\begin{equation}
    \varepsilon = \sqrt{\frac{1}{N_{\text{channels}}}\sum_{k=1}^{N_{\text{channels}}}\frac{\left[\left(\by_{21}-\bgamma_{21}\right)_k\right]^2}{(\bDelta_{21})_{kk}}}\,, \label{eq:signal-bias-statistic}
\end{equation}
where $\by_{21}$ is the input signal, and $\bgamma_{21}$ and $\bDelta_{21}$ are the mean and $1\sigma$ channel covariance calculated in Equations~\ref{eq:channel-mean}~and~\ref{eq:channel-covariance}.~Figure~\ref{fig:linear-calibration} shows the cumulative distribution functions of 5000 fits for the flattened Gaussian (blue curve) and for the turning point (orange curve) models and indicates that the $95\%$ confidence intervals for the flattened Gaussian (turning point) fits correspond to $8.75\sigma$ ($2.5\sigma$). This difference indicates that it is much easier to linearly separate the beam-weighted foreground model from the turning point model than from the flattened Gaussian model.

\begin{figure}[tb]
  \centering
  \includegraphics[width=0.4\textwidth]{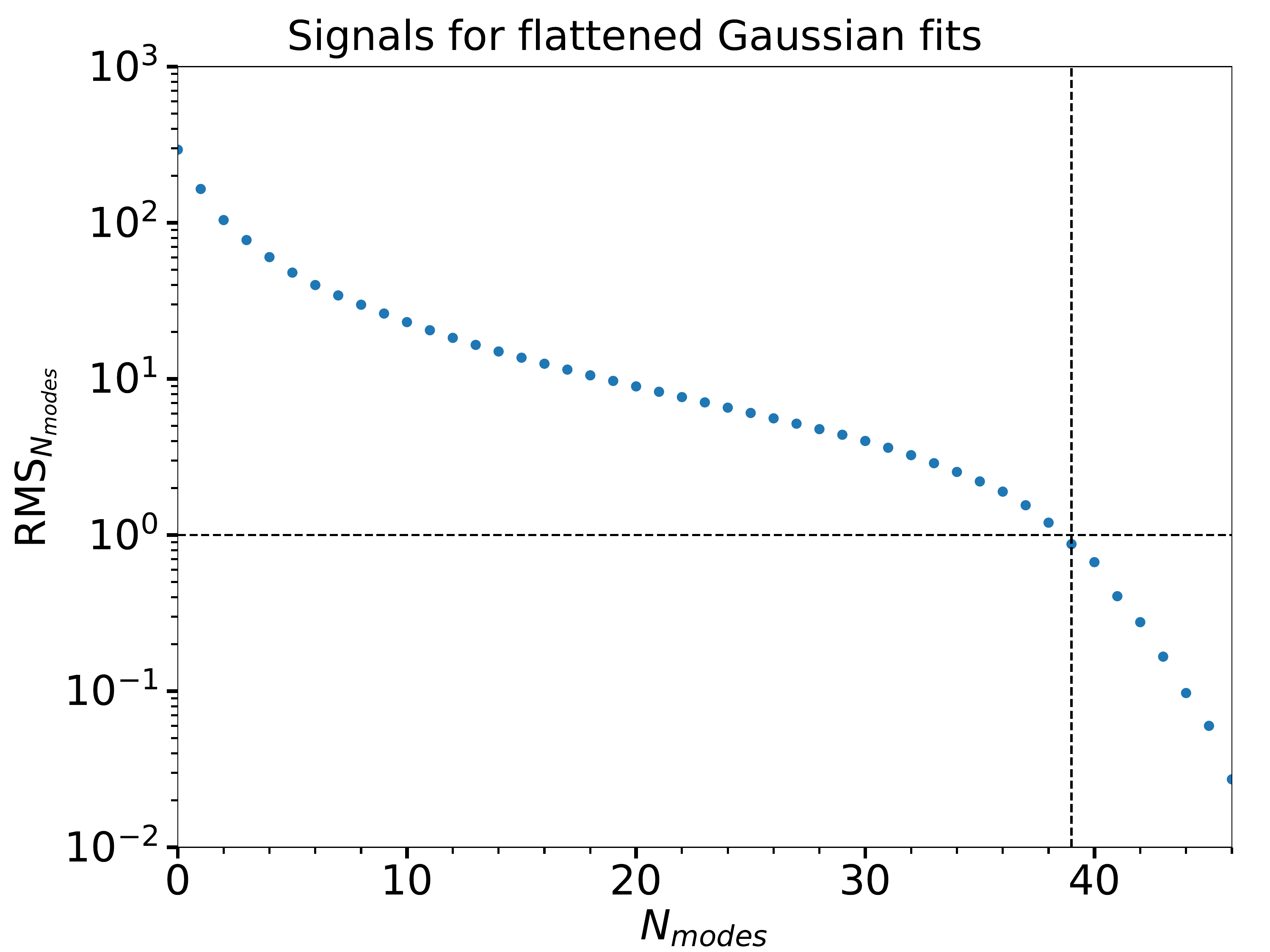}
  \includegraphics[width=0.4\textwidth]{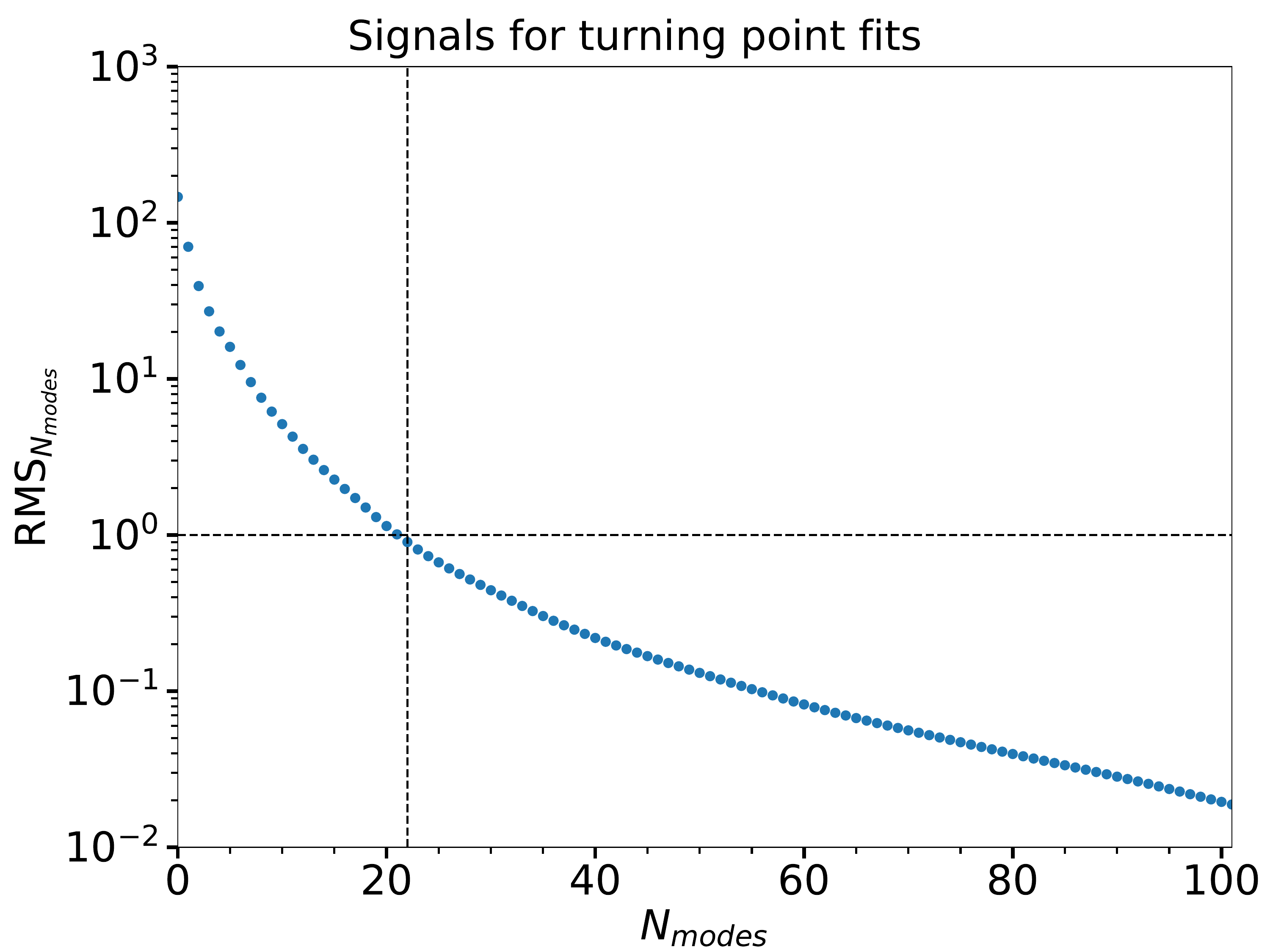}
  \includegraphics[width=0.4\textwidth]{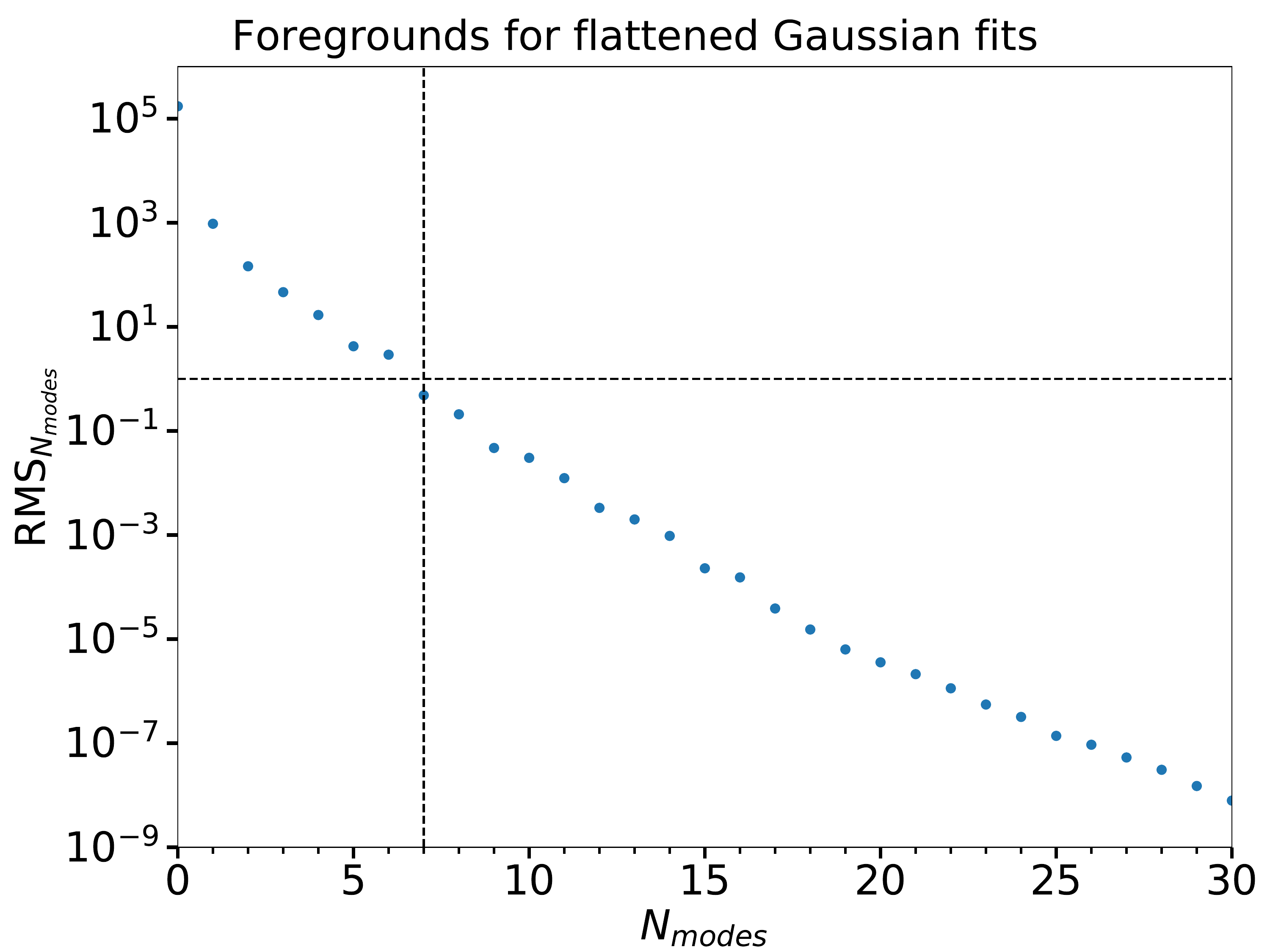}
  \includegraphics[width=0.4\textwidth]{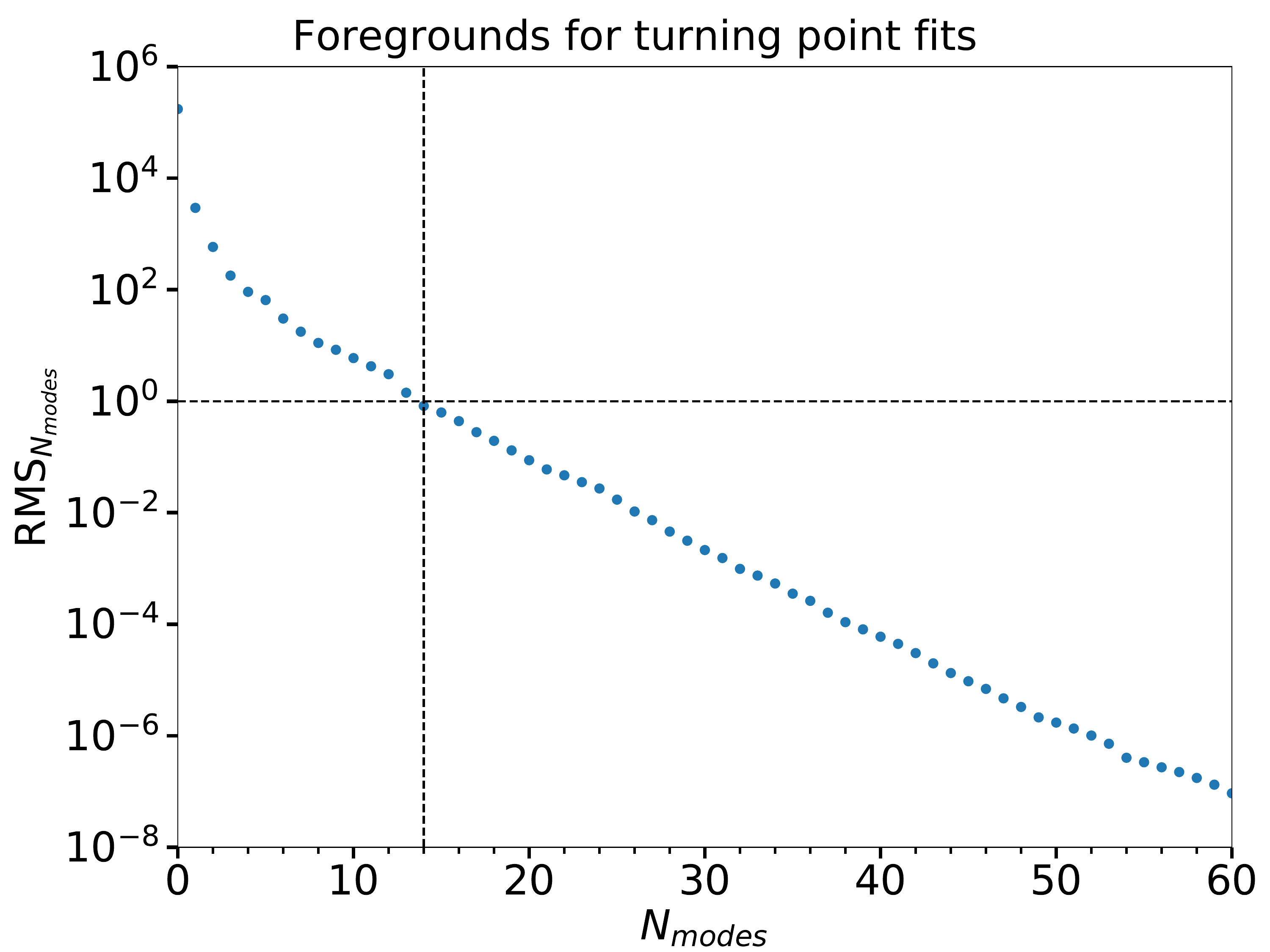}
  \caption{RMS error as a function of number of modes, $N_{modes}$, as given by Equation~\ref{eq:importance-spectrum} for four different training sets. The top, left (right) panel corresponds to the flattened Gaussian (turning point) model signal training set, while the bottom, left (right) panel corresponds to the foreground training set used for the flattened Gaussian (turning point) model fits. Values are given in terms of numbers of noise levels assuming 800 hours of integration. The horizontal, dashed lines mark $\text{RMS}_{N_{\text{modes}}}=1$ and the vertical, dashed lines mark the smallest $N_{\text{modes}}$ values where $\text{RMS}_{N_{\text{modes}}}<1$.~If at least these many modes are used when describing a curve from the training set, it is relatively likely that the curve can be fit to within the noise level.} \label{fig:signal-mode-importances}
\end{figure}

\section{Training set eigenvalue spectra} \label{app:eigenvalue-spectra}

The eigenmodes of each training set, $\bB$, are given by its singular value decomposition, $\bB=\bU\bSigma\bV^T$ where $\bU^T\bC^{-1}\bU=\bI$ ($\bC$ is the noise covariance matrix), $\bV^T\bV=\bI$, and $\bSigma$ has the same shape as $\bB$ ($N_{\text{channels}}\times N_{\text{curves}}$) but all off-diagonal elements are zero and the diagonal elements are non-negative and decreasing. The matrix $\bF_{N_{\text{modes}}}$, whose columns form the basis vectors of our model of the component described by the training set when we choose $N_{\text{modes}}$ modes, consists of the first $N_{\text{modes}}$ columns of $\bU$. The Root-Mean-Square (RMS) error in number of noise levels when $\bF_{N_{\text{modes}}}$ is used to fit the curves of the training set is
\begin{subequations} \begin{align}
    \text{RMS}_{N_{\text{modes}}} &= \sqrt{\frac{1}{N_{\text{channels}}\ N_{\text{curves}}}\text{ Tr}\left\{\left[\left(\bI-\bF_{N_{\text{modes}}}\bF_{N_{\text{modes}}}^T\bC^{-1}\right)\bB\right]^T\bC^{-1}\left[\left(\bI-\bF_{N_{\text{modes}}}\bF_{N_{\text{modes}}}^T\bC^{-1}\right)\bB\right]\right\}} \\
    &= \sqrt{\frac{1}{N_{\text{channels}}\ N_{\text{curves}}}\sum_{n=N_{\text{modes}}+1}^{\text{min}(N_{\text{channels}},N_{\text{curves}})}\sigma_{n}^2}\,, \label{eq:importance-spectrum}
\end{align} \end{subequations}
where $\sigma_n=\Sigma_{nn}$.~Figure~\ref{fig:signal-mode-importances} shows the importance spectrum of the flattened Gaussian and turning point models training sets using the $\text{RMS}_{N_\text{modes}}$ metric, assuming 800 hours of integration.

\section{Priors from SVD matrices} \label{app:priors-from-svd}

This appendix concerns prior distributions inferred from training sets.~As such, all fits discussed here refer to only one data component, taken here to be the foreground. If we choose orthonormal basis vectors for our foreground model (using $\bof_j^T\bC^{-1}\bof_k=\delta_{jk}$ as the orthonormality condition, where $\bof_k$ is the $k^{\text{th}}$ basis vector and $\bC$ is the noise covariance matrix), then the distribution of weights on the $n^{\text{th}}$ mode $\bof_n$ when fitting a training set $\bB$ is given by (see, e.g., Equation~\ref{eq:signal-projection-matrix})
\begin{equation}
    \bx_n = \bof_n^T\bC^{-1}\bB.
\end{equation}
Here we assume, as above (see Appendix~\ref{app:eigenvalue-spectra}), that $\bB=\bU\bSigma\bV^T$, where $\bU^T\bC^{-1}\bU=\bI$, $\bV^T\bV=\bI$, and $\bSigma$ is the same shape as $\bB$ but all off-diagonal elements are zero and the diagonal elements are non-negative and decreasing. Therefore,
\begin{equation}
    \bx_n = \bof_n^T\bC^{-1}\bU\bSigma\bV^T.
\end{equation}
Since we choose our basis vectors $\{\bof_k\}$ via the SVD of $\bB$, $\bof_n$ is the $n^{\text{th}}$ column of $\bU$. Because the columns of $\bU$ are orthonormal (automatically satisfying our orthonormality condition from earlier), $(\bof_n^T\bC^{-1}\bU)_k=\delta_{nk}$ where $\delta_{ij}$ is 1 if $i=j$ and 0 otherwise. Writing $\Sigma_{jk}=\sigma_j\delta_{jk}$ to account for its diagonal nature and defining $\bov_k$ as the $k^{\text{th}}$ column of $\bV$, it can be seen that $\bx_n=\sigma_n\bov_n$. Hence, if we define $\bj$ as a vector with the same dimension as $\bx_n$ (the number of training set vectors, $N$) whose elements are all 1, and note that $|\bov_n|^2=1$ since $\bV^T\bV=\bI$, then the mean and covariance of the $n^{\text{th}}$ mode weight can be written
\begin{subequations} \begin{align}
  \text{E}[x_n] &= \frac{\sigma_n}{N}\ \bj^T\bov_n, \\
  \text{Var}[x_n] &= \frac{{\sigma_n}^2}{N}\ \left[1 - \frac{(\bj^T\bov_n)^2}{N}\right].
\end{align} \end{subequations}
This information from the training set is used to seed Gaussian priors that allow us to suppress variations in unimportant modes (Section~\ref{sec:foregroundterms}). While in principle covariances of the mode weights in the training set could be added, this could lead to numerical issues when inverting the covariance matrix. In addition, it is conservative to use only the variances in the priors.

\section{Choosing a proposal covariance matrix from an estimated covariance matrix}
\label{app:updating}

In Sections~\ref{sec:proposal-distribution}~and~\ref{sec:updating} we use a function $c(\alpha)$ that is the constant of proportionality between the estimated covariance matrix of a distribution and the optimal proposal covariance matrix with which to explore that distribution given a desired acceptance fraction of proposals, $\alpha$. Assuming the distribution to explore is Gaussian with mean $\overline{\bx}$ and covariance $\bLambda$, and the proposal distribution is also Gaussian, the acceptance fraction when jumping from $\overline{\bx}$ if the proposal distribution has covariance $\bLambda/c$ is equal to
\begin{equation}
  \alpha=\left(1+\frac{1}{c}\right)^{-N/2},
\end{equation}
where $N$ is the dimension of the Gaussians. Solving for $c$ as a function of $\alpha$, this is
\begin{equation}
  c(\alpha) = \frac{1}{\alpha^{-2/N}-1}\,. \label{eq:proposal-proportionality}
\end{equation}
Thus, the Gaussian distribution which leads to an acceptance fraction of $\alpha$ when exploring a distribution with covariance $\bLambda$ from its mean is $\bLambda/c(\alpha)$.

\bibliographystyle{yahapj}
\bibliography{references}

\end{document}